\documentclass[letter]{article}

\usepackage{pdfpages}
\usepackage{fullpage}
\usepackage{textcomp}
\usepackage{setspace}
\usepackage{subcaption}
\usepackage{enumerate}
\usepackage{amsmath}
\usepackage{amsfonts}
\usepackage{graphicx}
\usepackage{amssymb}
\usepackage{multirow}
\usepackage{geometry}
\usepackage{soul}
\usepackage{colortbl}
\usepackage{wrapfig}

\definecolor{mygray}{gray}{.9}

\newtheorem{lemma}{Lemma}
\newtheorem{lemma*}[lemma]{Lemma*}
\numberwithin{lemma}{section}
\newtheorem{theorem}[lemma]{Theorem}
\newtheorem{theorem*}[lemma]{Theorem*}

\newtheorem{definition}[lemma]{Definition}

\usepackage{wrapfig}
\usepackage{bm}

\title{Colored range closest-pair problem under general distance functions}
\author{
  Jie Xue \\ University of Minnesota, Twin Cities \\ \texttt{xuexx193@umn.edu}
}
\date{}

\begin{document}

\maketitle

\begin{spacing}{1.0}

\begin{abstract}
The \textit{range closest-pair} (RCP) problem is the range-search version of the classical closest-pair problem, which aims to store a given dataset of points in some data structure such that when a query range $X$ is specified, the closest pair of points contained in $X$ can be reported efficiently.
A natural generalization of the RCP problem is the \textit{colored range closest-pair} (CRCP) problem in which the given data points are colored and the goal is to find the closest \textit{bichromatic} pair contained in the query range.
All the previous work on the RCP problem was restricted to the uncolored version and the Euclidean distance function.
In this paper, we make the first progress on the CRCP problem.
We investigate the problem under a general distance function induced by a monotone norm; in particular, this covers all the $L_p$-metrics for $p > 0$ and the $L_\infty$-metric.
We design efficient $(1+\varepsilon)$-approximate CRCP data structures for orthogonal queries in $\mathbb{R}^2$, where $\varepsilon>0$ is a pre-specified parameter.
The highlights are two data structures for answering rectangle queries, one of which uses $O(\varepsilon^{-1} n \log^4 n)$ space and $O(\log^4 n + \varepsilon^{-1} \log^3 n + \varepsilon^{-2} \log n)$ query time while the other uses $O(\varepsilon^{-1} n \log^3 n)$ space and $O(\log^5 n + \varepsilon^{-1} \log^4 n + \varepsilon^{-2} \log^2 n)$ query time.
In addition, we also apply our techniques to the CRCP problem in higher dimensions, obtaining efficient data structures for slab, 2-box, and 3D dominance queries.
Before this paper, almost all the existing results for the RCP problem were achieved in $\mathbb{R}^2$.
\end{abstract}

\section{Introduction}
The closest-pair problem, which aims to find the closest pair of points in a given dataset of points in $\mathbb{R}^2$ (or more generally $\mathbb{R}^d$), is one of the most fundamental problems in computational geometry and finds many applications in various scenarios, e.g., traffic control, similarity search, etc.
A natural and important generalization of this problem is the \textit{colored} closest-pair problem in which the given data points are colored and the point pairs of interest are only the \textit{bichromatic} ones (i.e., those consisting of two points of different colors), namely, we want to find the closest bichromatic pair of points.
This colored version has applications in analyzing categorical data, and is strongly related to Euclidean minimum spanning tree \cite{agarwal1991euclidean,eppstein1995dynamic}.
The closest-pair problem and its generalization can be considered under the Euclidean metric or more general metrics (such as $L_p$-metrics).

The \textit{range closest-pair} (RCP) problem, introduced in \cite{shan2003spatial} for the first time, is the range-search version of the classical (single-shot) closest-pair problem which aims to store a given dataset $S$ into some data structure which can report, for a specified query range $X$, the closest pair of points in $S \cap X$.
The RCP problem in $\mathbb{R}^2$ (under the Euclidean metric) has been studied in prior work over the last decades, see for example \cite{abam2009power,gupta2006range,gupta2014data,shan2003spatial,sharathkumar2007range,xue2018new}.

Compared to traditional range-search problems, the RCP problem (even the uncolored version) is much more challenging due to a couple of reasons.
First, as a range-search problem, the RCP problem is \textit{non-decomposable} in the sense that even if the dataset $S$ can be written as $S = S_1 \cup S_2$, the closest pair of points in $S \cap X$ cannot be obtained via those in $S_1 \cap X$ and $S_2 \cap X$.
This makes the decomposition-based techniques inapplicable to the RCP problem.
Second, since the RCP problem concerns the pairwise distances of the points, it is difficult to apply ``mapping''-based approaches to solve the problem.
For example, it is well-known that 2D circular range reporting can be reduced to 3D halfspace range reporting via a lifting argument.
However, this reduction does not work for the RCP problem because the lifting changes the pairwise distances of the points.
For another example, consider vertical strip queries in $\mathbb{R}^2$.
Range reporting for vertical strip queries is ``pointless'', as it is just the 1D range-reporting problem, after projecting the data points to the $x$-axis.
Again, this projection argument does not apply to the RCP problem as it changes the pairwise distances of the points, and the RCP problem for vertical strip queries is actually nontrivial \cite{sharathkumar2007range,xue2018new}.

Similarly to the single-shot closest-pair problem, the RCP problem can be naturally generalized to the \textit{colored range closest-pair} (CRCP) problem in which we want to store a colored dataset and report the closest bichromatic pair of points contained in a query range.
Surprisingly, despite of much effort made on the RCP problem, this generalization has never been considered previously.
In this paper, we make the first progress on the CRCP problem.
Unlike the previous work, we do not restrict ourselves to the Euclidean metric.
Instead, we investigate the problem under a general metric satisfying some certain condition.
This covers all $L_p$-metrics for $p>0$ (including the $L_\infty$-metric).
The CRCP problem is even harder than the (uncolored) RCP problem, especially when considered under such a general metric.
As such, we are interested in answering CRCP queries \textit{approximately}.
That is, for a specified query range $X$, we want to report a bichromatic pair of points in $X$ whose distance is at most $(1+\varepsilon) \cdot \mathsf{Opt}$ where $\mathsf{Opt}$ is the distance of the closest bichromatic pair of points in $X$, where $\varepsilon$ is a pre-specified parameter.
Our main goal is to design efficient $(1+\varepsilon)$-approximate CRCP data structures in terms of \textit{space} and \textit{query time}.
\smallskip


\noindent
\textbf{Related work.}
The closest-pair problem and range search are both the most fundamental problems in Computational Geometry, see \cite{agarwal1999geometric:range_search,smid1995closest} for surveys.
Approximate range search is also well-studied in the last decades, see for example \cite{arya2000approximate,chazelle2008approximate}.
The RCP problem was introduced by Shan et al. \cite{shan2003spatial} for the first time.
Subsequently, the problem in $\mathbb{R}^2$ was studied by \cite{abam2009power,gupta2006range,gupta2014data,sharathkumar2007range}.
The papers \cite{gupta2006range,gupta2014data,sharathkumar2007range} considered the problem with orthogonal queries, while \cite{abam2009power} studied halfplane queries.
Recently, Xue et al. \cite{xue2018new} improved the above results.
The state-of-the-art RCP data structure for rectangle query uses $O(n \log^2 n)$ space and $O(\log^2 n)$ query time \cite{xue2018new}.
In higher dimensions, the RCP problem is still open.
To our best knowledge, the only known result that can be generalized to higher dimensions is a simple data structure given in \cite{gupta2014data}, which only has guaranteed \textit{average-case} performance.
All these results were limited to the Euclidean metric and uncolored case.

\subsection{Our contributions and techniques} \label{sec-contribution}
In this paper, we investigate the CRCP problem under a general metric that is induced by a monotone norm (see Section~\ref{sec-notation} for the definition); in particular, this includes all $L_p$-metrics for $p>0$ and the $L_\infty$-metric.
We design $(1+\varepsilon)$-approximate CRCP data structures for orthogonal queries in $\mathbb{R}^2$ and higher dimensions, where $\varepsilon>0$ is a pre-specified parameter.
The performances of these data structures are summarized in Table~\ref{tab-result}, and we give a brief explanation below.
\begin{table}[h]
    \centering
    \begin{tabular}{|c|c|c|c|c|}
        \hline
        Dimension & Query & Source & Space & Query time  \\
        \hline
        \multirow{4}{*}{$\mathbb{R}^2$} & Strip & Theorem~\ref{thm-strip} & $O(\varepsilon^{-1} n \log^2 n)$ & $O(\log n + \log(1/\varepsilon))$ \\
        \cline{2-5}
        & Quadrant & Theorem~\ref{thm-quadrant} & $O(\varepsilon^{-1} n \log^2 n)$ & $O(\log n + \log(1/\varepsilon))$ \\
        \cline{2-5}
        & \cellcolor[gray]{0.9} & \cellcolor[gray]{0.9}Theorem~\ref{thm-rect1} & \cellcolor[gray]{0.9}$O(\varepsilon^{-1} n \log^4 n)$ & \cellcolor[gray]{0.9}$O(\log^4 n + \varepsilon^{-1}\log^3 n + \varepsilon^{-2}\log n)$ \\
        \cline{3-5}
        & \cellcolor[gray]{0.9}\multirow{-2}{*}{Rectangle} & \cellcolor[gray]{0.9}Theorem~\ref{thm-rect2} & \cellcolor[gray]{0.9}$O(\varepsilon^{-1} n \log^3 n)$ & \cellcolor[gray]{0.9}$O(\log^5 n + \varepsilon^{-1}\log^4 n + \varepsilon^{-2}\log^2 n)$ \\
        \hline
        \multirow{2}{*}{$\mathbb{R}^d$} & Slab & Theorem~\ref{thm-slab} & $O(\varepsilon^{-1} n \log^d n)$ & $O(\log n + \log(1/\varepsilon))$ \\
        \cline{2-5}
        & 2-Box & Theorem~\ref{thm-2box} & $O(\varepsilon^{-1} n \log^d n)$ & $O(\log n + \log(1/\varepsilon))$ \\
        \hline
        $\mathbb{R}^3$ & Dominance & Theorem~\ref{thm-3dom} & $O(\varepsilon^{-1} n \log^6 n)$ & $O(\varepsilon^{-2} \log^9 n + \varepsilon^{-4} \log^3 n)$ \\
        \hline
    \end{tabular}
    \caption{The performances of our $(1+\varepsilon)$-approximate CRCP data structures.}
    \label{tab-result}
\end{table}

Our main result is two $(1+\varepsilon)$-approximate CRCP data structures for rectangle queries in $\mathbb{R}^2$; see the gray rows of Table~\ref{tab-result}.
In the process of designing these data structures, we also obtain efficient data structures for strip and quadrant queries.
Using similar techniques, we also achieve results in higher dimensions.
Specifically, we design data structures for slab and 2-box queries in $\mathbb{R}^d$ (which are generalizations of strip and quadrant queries in $\mathbb{R}^2$ respectively) and dominance queries in $\mathbb{R}^3$.
All of our data structures use \textit{near-linear} space and \textit{poly-logarithmic} query time (when $\varepsilon$ is regarded as a constant).
Preprocessing time is not considered in this paper, and we leave this as an open question for future work\footnote{Preprocessing RCP-related data structures is usually a hard task. For instance, how to build efficiently the state-of-the-art orthogonal RCP data structures in \cite{xue2018new} is still unknown.}.
Our new results are interesting for the following reasons. 
\begin{itemize}
    \item Previously, only the uncolored RCP problem was studied.
    We make the first progress on the CRCP problem.
    Furthermore, we do not make any assumption on the coloring of the dataset.
    \item The previous work considered the RCP problem only under the Euclidean metric.
    Our results can be applied to a quite general class of metrics including all $L_p$-metrics.
    \item Almost all existing results on the RCP problem were restricted to $\mathbb{R}^2$.
    Our techniques give some results beyond that (while our main focus is still on $\mathbb{R}^2$), leading us towards better understanding of the CRCP problem in higher dimensions.
\end{itemize}

\begin{figure}[h]
    \centering
    \includegraphics[height=5.5cm]{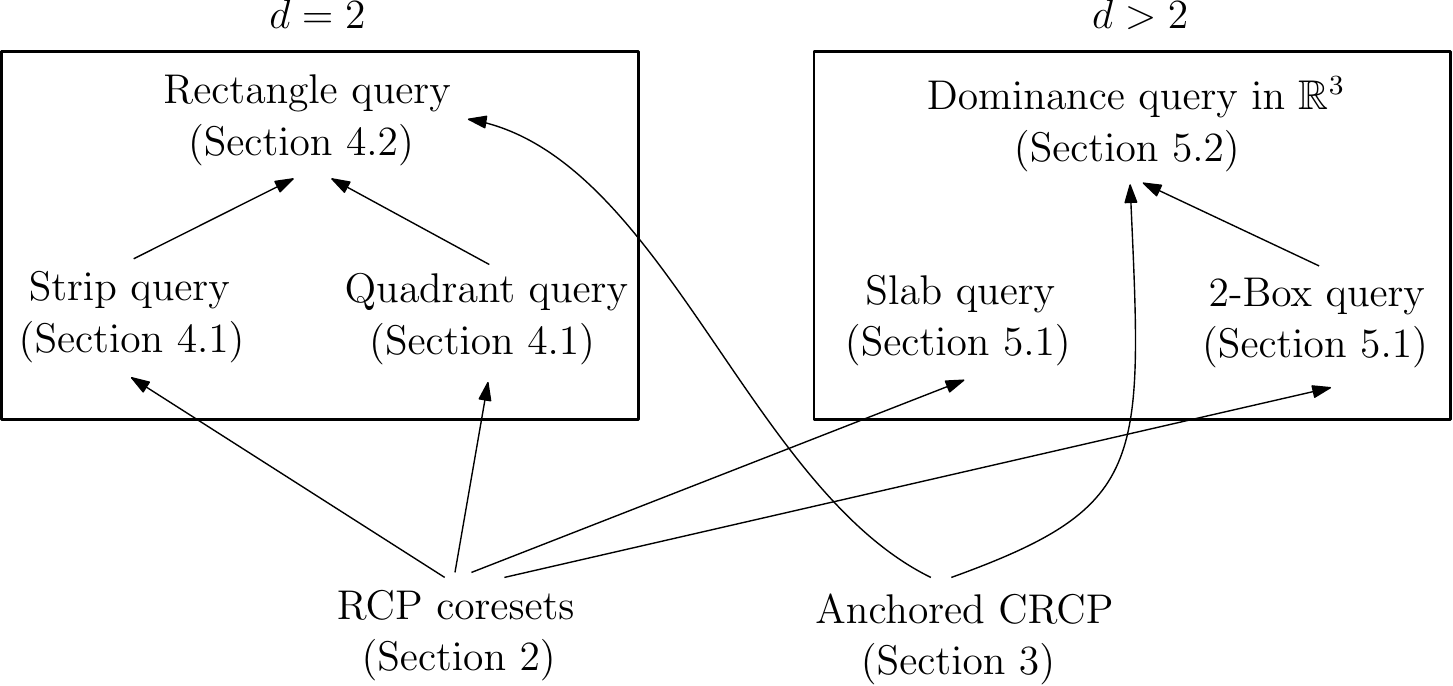}
    \caption{The dependency graph of our techniques and results}
    \label{fig-depend}
\end{figure}

\noindent
\textbf{Our techniques.}
Unfortunately, the techniques used in the (uncolored) RCP problem are inapplicable to the CRCP problem even under the Euclidean metric (we briefly argue this in Section~\ref{sec-difficult}).
Thus, we develop new techniques to solve the problem.
Our first technical contribution is the notion of \textit{RCP coresets}.
Roughly speaking, an RCP coreset of a set of point pairs is a subset that approximately preserves the closest-pair information in every query range.
This notion gives us a natural way to design an approximate CRCP data structure, namely, storing an RCP coreset of the set of all bichromatic pairs and searching for the answer in the coreset.
This idea works only when there exists a small-size RCP coreset.
We prove that if the query space (i.e., the collection of the query ranges) satisfies some nice property and the metric is induced by a monotone norm, then a small-size RCP coreset always exists.
Using this result, we obtain efficient approximate CRCP data structures for both strip and quadrant queries.
Our second technique is an \textit{anchored} version of the CRCP problem, in which an anchor point $o$ is specified with the query range and we want to report the closest $o$-anchored bichromatic pair contained in the query range (see Section~\ref{sec-anchored} for formal definitions).
We give an efficient (approximate) anchored CRCP data structure for rectangle queries, which works for any metric induced by a monotone norm.
Based on the above results, we design our rectangle CRCP data structures.
The main idea is to use range trees to reduce a rectangle CRCP query to several strip and quadrant CRCP queries and anchored CRCP queries.
Using our strip/quadrant CRCP data structures and anchored CRCP data structure mentioned above, we eventually obtain the two approximate CRCP data structures for rectangle queries.
Our results in higher dimensions are achieved using similar techniques and ideas.
See Figure~\ref{fig-depend} for a dependency graph of our results.
\smallskip

\noindent
\textbf{Organization.}
Section~\ref{sec-notation} presents the basic notions and preliminaries used throughout the paper.
We suggest the reader to read this section carefully before moving on.
Section~\ref{sec-coreset} and \ref{sec-anchored} discuss RCP coresets and anchored CRCP problem respectively, which are two technical cores of this paper.
Our results for the CRCP problem in $\mathbb{R}^2$ are given in Section~\ref{sec-2d}.
In Section~\ref{sec-higher}, we consider the CRCP problem in higher dimensions.
Finally, in Section~\ref{sec-difficult}, we give some evidences showing that designing (efficient) exact CRCP data structures might be difficult even for strip and quadrant queries.
To make the paper more readable, most proofs are deferred to Appendix~\ref{appx-proof}.
For convenience of the reader, we give proof sketches for some important lemmas in the main text.
The lemmas/theorems marked as ``*'' do not have proofs in Appendix~\ref{appx-proof}, as they are already clear from the arguments in the context.

\subsection{Notions and Preliminaries} \label{sec-notation}
We introduce the notions and preliminaries that are used throughout the paper.
\smallskip

\noindent
$\bullet$ \textbf{Basic notations.}
For a positive integer $n$, the notation $[n]$ denotes the set $\{1,\dots,n\}$.
For a point $a \in \mathbb{R}^2$, we use $\mathsf{x}(a)$ and $\mathsf{y}(a)$ to denote its $x$-coordinate and $y$-coordinate respectively.
For a set $A$ of points in $\mathbb{R}^d$, we denote by $\mathsf{BB}(A)$ the minimum (axis-parallel) bounding box of $A$.
\smallskip

\noindent
$\bullet$ \textbf{Norms and metrics.}
We say a norm $\lVert \cdot \rVert$ on $\mathbb{R}^d$ is \textit{monotone} provided that $\lVert (x_1,\dots,x_d) \rVert \geq \lVert (x_1',\dots,x_d') \rVert$ if $|x_i| \geq |x_i'|$ for all $i \in [d]$.
Note that all $L_p$-norms for $p>0$ and the $L_\infty$-norm are monotone, and a conical combination of monotone norms is also monotone.
A metric $\delta$ on $\mathbb{R}^d$ is induced by a norm $\lVert \cdot \rVert$ on $\mathbb{R}^d$ if $\delta(a,b) = \lVert a-b \rVert$ for all $a,b \in \mathbb{R}^d$.
A metric induced by a monotone norm has the following relation to the Euclidean metric $L_2(\cdot,\cdot)$.
\begin{lemma} \label{lem-mono}
If $\delta$ is a metric on $\mathbb{R}^d$ induced by a monotone norm $\lVert \cdot \rVert$ and $\{e_1,\dots,e_d\}$ is the standard basis of $\mathbb{R}^d$, then $(1/\sqrt{d}) \cdot L_2(a,b) \cdot \min_{i \in [d]} \lVert e_i \rVert \leq \delta(a,b) \leq d \cdot L_2(a,b) \cdot \max_{i \in [d]} \lVert e_i \rVert$ for all $a,b \in \mathbb{R}^d$.
\end{lemma}

\noindent
$\bullet$ \textbf{Point pairs.}
A point pair $\phi$ in $\mathbb{R}^2$ is \textit{NE-SW} (resp., \textit{NW-SE}) if the two points of $\phi$ are the northeast and southwest (northwest and southeast) vertices of $\mathsf{BB}(\phi)$ respectively\footnote{If the two points of $\phi$ have the same $x$-coordinates or $y$-coordinates, then $\phi$ is both NE-SW and NW-SE.}.
Let $\varPi$ be a set of point pairs in $\mathbb{R}^d$.
We define the \textit{ground set} of $\varPi$ as $\mathsf{Grd}(\varPi) = \bigcup_{\phi \in \varPi} \phi$.
Note that $\mathsf{Grd}(\varPi)$ is \textit{not} a multi-set, thus it can happen $|\mathsf{Grd}(\varPi)|<2|\varPi|$.
For a range $X$ in $\mathbb{R}^d$, we denote by $\varPi \Cap X$ the subset of $\varPi$ consisting of all pairs contained in $X$, i.e., $\varPi \Cap X = \{\phi \in \varPi: \phi \subseteq X\}$.
\smallskip

\noindent
$\bullet$ \textbf{Colored datasets.}
Let $S$ be a set of colored points in $\mathbb{R}^d$.
For a point $a \in S$, we use $\text{cl}(a)$ to denote the color of $a$.
A pair $\phi = \{a,b\}$ of points in $S$ is called \textit{bichromatic} if $\text{cl}(a) \neq \text{cl}(b)$.
We use the notation $\mathsf{Bich}(S)$ to denote the set of all the bichromatic pairs of points in $S$.
\smallskip

\noindent
$\bullet$ \textbf{Closest pair and approximate closest pair (ACP).}
Let $\delta$ be a metric on $\mathbb{R}^d$.
For a point pair $\phi = \{a,b\}$ in $\mathbb{R}^d$, we define $|\phi|_\delta = \delta(a,b)$.
Let $\varPi$ be a set of point pairs in $\mathbb{R}^d$.
We denote by $\mathsf{CP}_\delta(\varPi)$ the closest pair in $\varPi$ under the metric $\delta$, i.e., $\mathsf{CP}_\delta(\varPi) = \arg\min_{\phi \in \varPi} |\phi|_\delta$.
A pair $\phi \in \varPi$ is a $(1+\varepsilon)$-\textit{approximate closest pair} (ACP) in $\varPi$ under the metric $\delta$ if $|\phi|_\delta \leq (1+\varepsilon) \cdot |\mathsf{CP}_\delta(\varPi)|_\delta$.
\smallskip

\noindent
$\bullet$ \textbf{Approximate CRCP data structures.}
Let $\mathcal{X}$ be a collection of ranges in $\mathbb{R}^d$ and $\delta$ be a metric on $\mathbb{R}^d$.
For a parameter $\varepsilon > 0$, a $(1+\varepsilon)$-\textit{approximate} $(\mathcal{X},\delta)$-\textit{CRCP data structure} built on a colored dataset $S$ in $\mathbb{R}^d$ can return, for any specified $X \in \mathcal{X}$, a $(1+\varepsilon)$-ACP in $\mathsf{Bich}(S) \Cap X$.

\section{RCP coresets} \label{sec-coreset}
Coresets are commonly used in Computational Geometry for designing approximation geometric algorithms \cite{agarwal2005geometric,frahling2005coresets,har2006coresets,har2004coresets}.
A coreset is (roughly) a small subset that approximately captures some geometric measure of the entire set.
In this section, we define coresets in the context of RCP-related problems, which will play an important role in our approximate CRCP data structures.

To get some intuition, let us consider the CRCP problem.
In the CRCP problem, the objects of interest are bichromatic pairs in the given colored dataset.
However, the number of these pairs can be quadratic.
So we ask the following question in the spirit of coresets: does there exist a small subset of these pairs that approximately captures the closest-pair information in every query range?
This results in the notion of \textit{RCP coresets} formally defined below.
Let $\varPi$ be a set of point pairs in $\mathbb{R}^d$, $\mathcal{X}$ be a collection of ranges in $\mathbb{R}^d$, and $\delta$ be a metric on $\mathbb{R}^d$.
\begin{definition}
A subset $\varPi' \subseteq \varPi$ is called an $\varepsilon$-\textbf{RCP coreset} \textnormal{(}or $\varepsilon$-\textbf{coreset} for short\textnormal{)} of $\varPi$ for $\mathcal{X}$-queries under the metric $\delta$ provided that for all $X \in \mathcal{X}$ with $\varPi \Cap X \neq \emptyset$, we have $\varPi' \Cap X \neq \emptyset$ and $|\mathsf{CP}_\delta(\varPi' \Cap X)|_\delta \leq (1+\varepsilon) \cdot |\mathsf{CP}_\delta(\varPi \Cap X)|_\delta$.
In the context where $\mathcal{X}$ and $\delta$ are both clear, we simply say $\varPi'$ is an $\varepsilon$-coreset of $\varPi$.
\end{definition}

The notion of RCP coreset gives us a natural way to design a $(1+\varepsilon)$-approximate CRCP data structure.
Specifically, we can store an $\varepsilon$-coreset $\varPi'$ of $\varPi = \mathsf{Bich}(S)$ and, when a query range $X$ is specified, report $\mathsf{CP}_\delta(\varPi' \Cap X)$ as the answer, which is obviously a $(1+\varepsilon)$-ACP in $\mathsf{Bich}(S) \Cap X$.
Clearly, this idea works only when there exists such an $\varepsilon$-coreset of a small size.
Therefore, in the next section, we shall study the existence of small-size coresets.

\subsection{Small-size coresets for well-behaved query spaces}
In this section, we show that any finite set of point pairs has a small-size $\varepsilon$-coreset as long as the query space $\mathcal{X}$ is ``well-behaved'' and the metric $\delta$ is induced by a monotone norm.
\begin{definition}
Two point pairs $\phi$ and $\psi$ in $\mathbb{R}^d$ are \textbf{adjacent} if they share a common point, i.e., $|\phi \cap \psi| = 1$, and are \textbf{strongly adjacent} if in addition the common point is a vertex of $\mathsf{BB}(\phi \cup \psi)$.
\end{definition}
\begin{definition} \label{def-well}
Let $\varPi$ be a set of point pairs in $\mathbb{R}^d$, and $\mathcal{X}$ be a collection of ranges in $\mathbb{R}^d$.
We say $\mathcal{X}$ is \textbf{well-behaved} on $\varPi$ if it satisfies the following two conditions: \\
$\bullet$ For each pair $\phi \in \varPi$, there exists a ``smallest'' element $X_\phi \in \mathcal{X}$ containing $\phi$ in the sense that $X_\phi \subseteq X$ for all $X \in \mathcal{X}$ containing $\phi$. \\
$\bullet$ For any two strongly adjacent pairs $\phi,\psi \in \varPi$, either $X_\phi \subseteq X_\psi$ or $X_\psi \subseteq X_\phi$.
\end{definition}
For example, the collection of vertical strips is well-behaved on any set of point pairs in $\mathbb{R}^2$ and the collection of northeast quadrants is well-behaved on any set of NW-SE pairs in $\mathbb{R}^2$; see Appendix~\ref{appx-well}.
We notice the following property of a well-behaved query space.
\begin{lemma} \label{lem-minimal}
If $\varPi \neq \emptyset$ is finite and $\mathcal{X}$ is well-behaved on $\varPi$, then there exists a minimal element $X^* \in \mathcal{X}$ satisfying $\varPi \Cap X^* \neq \emptyset$ \textnormal{(}the minimality is in terms of the partial order ``$\subseteq$'' \textnormal{)}.
\end{lemma}

Let $\varPi$ be a set of point pairs in $\mathbb{R}^d$, $\delta$ be a metric on $\mathbb{R}^d$ induced by a monotone norm, and $\mathcal{X}$ be a collection of ranges in $\mathbb{R}^d$ well-behaved on $\varPi$.
The rest of this section is dedicated to prove the existence of a small-size $\varepsilon$-coreset $\varPi'$ of $\varPi$ for $\mathcal{X}$-queries under the metric $\delta$.
Our proof is constructive.
We generate $\varPi'$ via the following simple procedure.
Initially, set $\varPi'=\emptyset$.
\begin{enumerate}
    \item Select a minimal element $X^* \in \mathcal{X}$ satisfying $\varPi \Cap X^* \neq \emptyset$ (Lemma~\ref{lem-minimal}).
    \item $\phi^* \leftarrow \mathsf{CP}_\delta(\varPi \Cap X^*)$.
    If $\varPi' \Cap X^*=\emptyset$ or $(1+\varepsilon) \cdot |\phi^*|_\delta < |\mathsf{CP}_\delta(\varPi' \Cap X^*)|_\delta$, then $\varPi' \leftarrow \varPi' \cup \{\phi^*\}$.
    \item $\varPi \leftarrow \varPi \backslash \{\phi^*\}$.
    If $\varPi \neq \emptyset$, go back to Step 1; otherwise, return the set $\varPi'$.
\end{enumerate}
It is not difficult to verify the correctness of the above procedure.
\begin{lemma} \label{lem-corecorrect}
The set $\varPi'$ returned by the above procedure is an $\varepsilon$-coreset of $\varPi$.
\end{lemma}

The difficult part is to bound the size of $\varPi'$ returned in the above procedure.
Let $n = |\mathsf{Grd}(\varPi)|$.
Our conclusion is that $|\varPi'| = O(\varepsilon^{-1} n \log^d n)$.
For simplicity of exposition, here we only prove this bound for $d=2$.
The generalization of the proof to higher dimensions is somehow straightforward, and is deferred to the complete proof of Theorem~\ref{thm-coreset}.
We begin with introducing some notations.
For $m \in [n]$, we denote by $\mathcal{G}_m$ the collection of all subsets of $\mathsf{Grd}(\varPi)$ of size $m$.
For $G \subseteq \mathsf{Grd}(\varPi)$, we write $\varPi'_G = \varPi' \Cap G$, which is by definition the subset of $\varPi'$ consisting of all pairs with two points in $G$. 
Define a function $f:[n] \rightarrow \mathbb{N}$ as $f(m) = \max_{G \in \mathcal{G}_m} |\varPi'_G|$.
Note that $f(n) = |\varPi'|$.

Our goal is to show that $f(m) = O(\varepsilon^{-1} m \log^2 m)$.
To this end, we first try to write $f$ in a recursive form.
Fix $m \in [n]$ and $G \in \mathcal{G}_m$.
Let $G_1 \subseteq G$ (resp., $G_2 \subseteq G$) consist of the left (resp., right) $m/2$ points in $G$, and $v$ be a vertical line separating $G_1$ and $G_2$ (i.e., with $G_1$ on the left side and $G_2$ on the right side).
Now we have $\varPi'_G = \varPi'_{G_1} \cup \varPi'_{G_2} \cup \varPi'_{G,v}$ where $\varPi'_{G,v} \subseteq \varPi'_G$ consists of the pairs that ``cross'' $v$ (i.e., whose two points lie on different sides of $v$).
Note that $|\varPi'_{G_i}| \leq f(m/2)$ for $i \in \{1,2\}$, which implies $|\varPi'_G| \leq 2f(m/2) + \varPi'_{G,v}$.
Since this holds for all $G \in \mathcal{G}_m$, we have the recurrence $f(m) \leq 2f(m/2)+f'(m)$ where $f'(m) = \sup_{G \in \mathcal{G}_m, v \in V} |\varPi'_{G,v}|$ (here $V$ denotes the set of all vertical lines in $\mathbb{R}^2$).
It suffices to bound the function $f'$.
Using the same argument as above (but separating a set $G$ horizontally), we can obtain a similar recurrence $f'(m) \leq 2f'(m/2)+f''(m)$ where $f''(m) = \sup_{G \in \mathcal{G}_m, v \in V, h \in H} |\varPi'_{G,v,h}|$ (here $H$ denotes the set of all horizontal lines in $\mathbb{R}^2$ and $\varPi'_{G,v,h} \subseteq \varPi'_G$ consists of the pairs that cross both $v$ and $h$).

The remaining task is to bound $f''(m)$, which is the trickiest part of the proof.
We shall show that $f''(m) \leq \lambda m$ where $\lambda = \lceil \log_{1+\varepsilon} 2 \rceil$ for all $m \in [n]$.
Fix $m \in [n]$.
By the definition of $f''(m)$, it suffices to show that $|\varPi'_{G,v,h}| \leq \lambda m$ for all $G \in \mathcal{G}_m$, $v \in V$, $h \in H$.
Let $G \in \mathcal{G}_m$ be an arbitrary element, and $v \in V$ (resp., $h \in H$) be an arbitrary vertical (resp., horizontal) line.
We prove $|\varPi'_{G,v,h}| \leq \lambda m$ by a charging argument.
For each pair $\phi \in \varPi'_{G,v,h}$, we charge it to one of its two points using the following rule.
Suppose $\phi = \{a,b\}$.
Define $\alpha$ (resp., $\beta$) as the number of the pairs $\psi \in \varPi'_{G,v,h}$ satisfying that \textbf{(i)} $a \in \psi$ (resp., $b \in \psi$) and \textbf{(ii)} $|\psi|_\delta < |\phi|_\delta$.
If $\alpha<\beta$ (resp., $\beta<\alpha$), we charge $\phi$ to $a$ (resp., $b$).
In the case $\alpha = \beta$, we arbitrarily charge $\phi$ to $a$ or $b$.
Our claim is that each point in $G$ is charged at most $\lambda$ times.
To see this, we first give some observations.
\begin{lemma} \label{lem-adjacent}
Two pairs in $\varPi'_{G,v,h}$ are strongly adjacent if they are adjacent.
\end{lemma}
\begin{lemma} \label{lem-gap}
If $\phi,\psi \in \varPi'$ are strongly adjacent, either $|\phi|_\delta > (1+\varepsilon) \cdot |\psi|_\delta$ or $|\psi|_\delta > (1+\varepsilon) \cdot |\phi|_\delta$.
\end{lemma}
\textit{Proof sketch.}
Assume $|\phi|_\delta < |\psi|_\delta$.
Since $\mathcal{X}$ is well-behaved on $\varPi$, either $X_\phi \subseteq X_\psi$ or $X_\psi \subseteq X_\phi$ by definition.
We observe that $X_\phi \nsubseteq X_\psi$, basically because if $X_\phi \subseteq X_\psi$, then $\phi$ is added to $\varPi'$ before $\psi$ in the procedure and $\psi$ cannot be added to $\varPi'$.
Thus, $X_\psi \subsetneq X_\phi$.
Based on this fact, we prove the lemma roughly as follows.
We first argue that $\psi$ is added to $\varPi'$ before $\phi$.
If $|\psi|_\delta \leq (1+\varepsilon) \cdot |\phi|_\delta$, then when $\phi$ is considered in the procedure, the existence of $\psi$ in $\varPi'$ prevents $\phi$ from being added to $\varPi'$.
As such, $|\psi|_\delta > (1+\varepsilon) \cdot |\phi|_\delta$.
\hfill $\Box$
\smallskip

\noindent
Assume a point $a \in G$ is charged more than $\lambda$ times.
Let $\phi_0,\phi_1,\dots,\phi_\lambda \in \varPi'_{G,v,h}$ be (any) $\lambda+1$ pairs charged to $a$, where $|\phi_0|_\delta < |\phi_1|_\delta < \cdots < |\phi_\lambda|_\delta$.
Suppose $\phi_\lambda = \{a,b\}$, and we denote by $\alpha$ (resp., $\beta$) the number of the pairs $\psi \in \varPi'_{G,v,h}$ satisfying that \textbf{(i)} $a \in \psi$ (resp., $b \in \psi$) and \textbf{(ii)} $|\psi|_\delta < |\phi_\lambda|_\delta$.
Note that $\alpha \geq \lambda$ because $\phi_0,\dots,\phi_{\lambda-1}$ all satisfy the two conditions above.
Since $\phi_\lambda$ is charged to $a$, we have $\alpha \leq \beta$ and thus $\beta \geq \lambda$.
As such, we can find $\lambda$ pairs $\psi_0,\dots,\psi_{\lambda-1} \in \varPi'_{G,v,h}$ such that $b \in \psi_i$ and $|\psi_i|_\delta < |\phi_\lambda|_\delta$ for all $i \in [\lambda-1]$.
Suppose $|\psi_0|_\delta < \cdots < |\psi_{\lambda-1}|_\delta$.
Since $\phi_0,\phi_1,\dots,\phi_\lambda$ are pairwise adjacent (they share the common point $a$), they are also pairwise strongly adjacent by Lemma~\ref{lem-adjacent}.
Then by Lemma~\ref{lem-gap}, we have $|\phi_i|_\delta > (1+\varepsilon)^i |\phi_0|_\delta$ for all $i \in [\lambda]$.
In particular, $|\phi_\lambda|_\delta > (1+\varepsilon)^\lambda |\phi_0|_\delta \geq 2 |\phi_0|_\delta$.
Using the same argument, we can also deduce that $|\phi_\lambda|_\delta > 2 |\psi_0|_\delta$.
Combining the two inequalities gives us $2|\phi_\lambda|_\delta > 2 |\phi_0|_\delta + 2 |\psi_0|_\delta$.
Therefore, $|\phi_\lambda|_\delta > |\phi_0|_\delta + |\psi_0|_\delta$.
Let $o$ be the intersection point of the lines $v$ and $h$.
Since $\delta$ is induced by a monotone norm, $|\phi_0|_\delta \geq \delta(a,o)$; indeed, if $\phi_0=\{a,c\}$, then $|\mathsf{x}(a)-\mathsf{x}(c)| \geq |\mathsf{x}(a)-\mathsf{x}(o)|$ and $|\mathsf{y}(a)-\mathsf{y}(c)| \geq |\mathsf{y}(a)-\mathsf{y}(o)|$ because $\phi_0$ crosses both $v$ and $h$.
Similarly, $|\psi_0|_\delta \geq \delta(b,o)$.
Now we have 
\begin{equation*}
\delta(a,b) = |\phi_\lambda|_\delta > |\phi_0|_\delta + |\psi_0|_\delta \geq \delta(a,o) + \delta(b,o).    
\end{equation*}
However, this violates the triangle inequality for the metric $\delta$.
As a result, each point in $G$ is charged at most $\lambda$ times and thus $|\varPi'_{G,v,h}| \leq \lambda m$.
This immediately shows that $f''(m) \leq \lambda m$.

With this in hand, the remaining proof is straightforward.
Since $f'(m) \leq 2f'(m/2) + f''(m)$, we have $f'(m) = O(\lambda m \log m)$ by Master Theorem.
Using Master theorem again to solve the recurrence $f(m) \leq 2f(m/2) + f'(m)$, we deduce $f(m) = O(\lambda m \log^2 m)$.
Therefore, $|\varPi'| = O(\lambda n \log^2 n)$.
Since $\lambda = \lceil \log_{1+\varepsilon} 2 \rceil = O(\varepsilon^{-1})$, we finally conclude $|\varPi'| = O(\varepsilon^{-1} n \log^2 n)$.

The above proof can be generalized to higher dimensions, resulting in the following theorem.
\begin{theorem} \label{thm-coreset}
Let $\varPi$ be a set of point pairs in $\mathbb{R}^d$, $\delta$ be a metric on $\mathbb{R}^d$ induced by a monotone norm, and $\mathcal{X}$ be a collection of ranges in $\mathbb{R}^d$ well-behaved on $\varPi$.
There exists an $\varepsilon$-coreset $\varPi'$ of $\varPi$ for $\mathcal{X}$-queries under the metric $\delta$ such that $|\varPi'| = O(\varepsilon^{-1} n \log^d n)$ where $n = |\mathsf{Grd}(\varPi)|$.
\end{theorem}

\section{Anchored CRCP data structures} \label{sec-anchored}
In this section, we define an \textit{anchored} variant of the CRCP problem and design efficient data structures for it, which is another technical ingredient of our solutions for the CRCP problem.

A pair $\phi$ of points in $\mathbb{R}^d$ is called $o$-\textit{anchored} for a point $o \in \mathbb{R}^d$ if $o \in \mathsf{BB}(\phi)$.
In the anchored CRCP problem, a query $(X,o)$ consists of a query range $X$ and an anchor point $o \in \mathbb{R}^d$, and we want to report the closest $o$-anchored bichromatic pair contained in $X$.
In other words, the only difference of the anchored CRCP problem (compared to the standard CRCP problem) is that an anchor point $o$ is specified in the query and we are only interested in the $o$-anchored pairs.
One can also consider the approximate version of the anchored CRCP problem, which aims to find an $o$-anchored bichromatic pair contained in $X$ that is a $(1+\varepsilon)$-approximation of the true answer.

In what follows, we consider the (approximate) anchored CRCP problem for rectangle query in $\mathbb{R}^2$ under a metric induced by a monotone norm.
We shall see how the anchor point specified in the query makes the problem ``easier'' than the standard CRCP problem.
Let $S$ be a colored dataset in $\mathbb{R}^2$, $\mathcal{R}$ be the collection of (axis-parallel) rectangles in $\mathbb{R}^2$, and $\delta$ be a metric on $\mathbb{R}^2$ induced by a monotone norm $\lVert \cdot \rVert$.
We assume $\lVert (1,0) \rVert = \lVert (0,1) \rVert = 1$, without loss of generality\footnote{Indeed, if this is not the case, we can define a one-to-one map $\pi:(x,y) \mapsto (x \cdot \lVert (1,0) \rVert,y \cdot \lVert (0,1) \rVert)$ on $\mathbb{R}^2$. Then the (anchored) CRCP problem on $S$ under the metric $\delta$ is equivalent to that on $\pi(S)$ under the metric $\delta'$ defined as $\delta'(a,b) = \delta(\pi^{-1}(a),\pi^{-1}(b))$. Note that $\delta'$ is also introduced by a monotone norm $\lVert \cdot \rVert'$ with $\lVert (1,0) \rVert' = \lVert (0,1) \rVert' = 1$.}.
For a parameter $\varepsilon>0$, our goal is to design a $(1+\varepsilon)$-approxmate anchored $(\mathcal{R},\delta)$-CRCP data structure.
To this end, consider a query $(R,o)$ where $R \in \mathcal{R}$ is a rectangle and $o \in \mathbb{R}^2$.
To answer the query, we need to report a $(1+\varepsilon)$-ACP in $\mathsf{Bich}_o(S) \Cap R$ under the metric $\delta$, where $\mathsf{Bich}_o(S) \subseteq \mathsf{Bich}(S)$ is the subset consisting of the pairs that are $o$-anchored.
Note that if $o \notin R$, then $\mathsf{Bich}_o(S) \Cap R = \emptyset$ and nothing should be reported.
So suppose $o \in R$.
We write $\mathsf{Bich}_o(S) = \varPi_1 \cup \varPi_2$ where $\varPi_1$ (resp., $\varPi_2$) consists of all NE-SW (resp., NW-SE) pairs in $\mathsf{Bich}_o(S)$.
Our data structure shall compute a $(1+\varepsilon)$-ACP $\phi_1$ (resp., $\phi_2$) in $\varPi_1 \Cap R$ (resp., $\varPi_2 \Cap R$), and then return $\phi^* = \mathsf{CP}_\delta(\{\phi_1,\phi_2\})$ as the answer, which is clearly a $(1+\varepsilon)$-ACP in $\mathsf{Bich}_o(S) \Cap R$.
It suffices to consider how to compute $\phi_1$. 

Let $v$ and $h$ denote the vertical and horizontal line through $o$, respectively.
Then $v$ and $h$ decompose $R$ into four small rectangles; we denote by $R^\text{ne}$ (resp., $R^\text{sw}$) the small rectangle to the northeast (resp., southwest) of $o$.
By definition, $\varPi_1 \Cap R$ is just the set of the bichromatic pairs with one point in $R^\text{ne}$ and one point in $R^\text{sw}$.
In order to find a $(1+\varepsilon)$-ACP $\phi_1$ in $\varPi_1 \Cap R$ under the metric $\delta$, we make the following key observation.
Set $\theta = \varepsilon/8$.

\begin{lemma} \label{lem-angle}
Let $a^*,a \in R^\textnormal{ne}$ such that $\angle a^*oa \leq \theta$ and $L_1(a^*,o) \leq L_1(a,o)$.
Let $b^*,b \in R^\textnormal{sw}$ such that $\angle b^*ob \leq \theta$ and $L_1(b^*,o) \leq L_1(b,o)$.
Then $\delta(a^*,b^*) \leq (1+\varepsilon) \cdot \delta(a,b)$.
\end{lemma}
\textit{Proof sketch.}
We can find a point $a'$ (resp., $b'$) on the segment connecting $o$ and $a$ (resp., $b$) such that $L_1(a^*,o) = L_1(a',o)$ (resp.,  $L_1(b^*,o) = L_1(b',o)$).
By the monotonicity of $\lVert \cdot \rVert$, we have $\delta(a',b') \leq \delta(a,b)$.
So it suffices to show $\delta(a^*,b^*) \leq (1+\varepsilon) \cdot \delta(a',b')$.
We have $\angle a^*oa' \leq \theta$ and $\angle oa^*a' \in [\pi/4,3\pi/4]$.
Thus, by sine law, $L_2(a^*,a')/L_2(a',o) \leq \sqrt{2} \theta$.
Applying Lemma~\ref{lem-mono} and the monotonicity of $\lVert \cdot \rVert$, we can deduce $\delta(a^*,a') \leq (\varepsilon/2) \cdot \delta(a',o) \leq (\varepsilon/2) \cdot \delta(a',b')$.
Similarly, $\delta(b^*,b') \leq (\varepsilon/2) \cdot \delta(a',b')$.
Then by the triangle inequality, $\delta(a^*,b^*) \leq (1+\varepsilon) \cdot \delta(a',b')$.
\hfill $\Box$
\smallskip

\noindent
\begin{wrapfigure}{r}{5.8cm}
    \vspace{-0.6cm}
	\begin{center}
		\includegraphics[height=4.5cm]{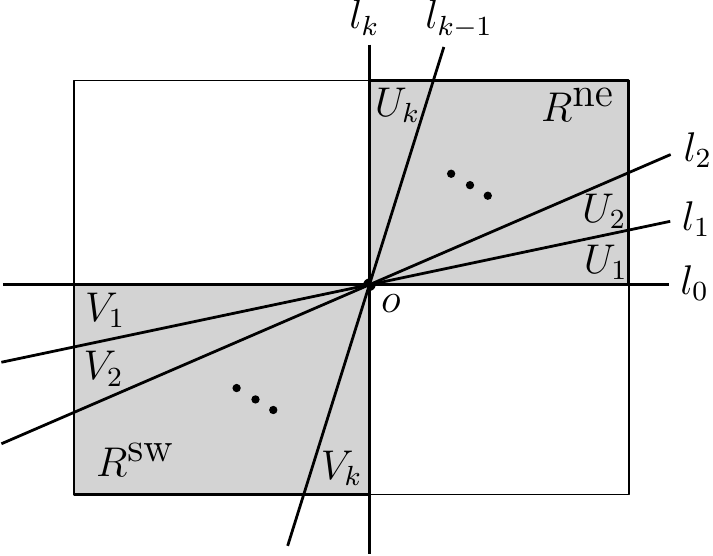}
	\end{center}
	\vspace{-0.6cm}
	\label{fig-angles}
\end{wrapfigure}
The above lemma holds even if we replace $L_1(\cdot,\cdot)$ with $L_2(\cdot,\cdot)$ (via a similar proof); we choose the $L_1$-metric for technical reasons, which will be clear later.
Now we use the above lemma to compute $\phi_1$.
Set $k = \lceil \pi/(2\theta) \rceil$.
Define $k+1$ lines $l_0,\dots,l_k$ as $l_i: (x-\mathsf{x}(o)) \cdot \cos(i\pi/2k) + (y-\mathsf{y}(o)) \cdot \sin(i\pi/2k) = 0$.
These lines go through $o$ and decompose $R^\text{ne}$ into $k$ regions, of which one is a 4-gon (the one containing the northeast vertex of $R^\text{ne}$) and the others are triangles.
We denote these regions by $U_1,\dots,U_k$ where $U_i$ is in between $l_{i-1}$ and $l_i$.
Note that $\angle a o a' \leq \theta$ for any $a,a' \in U_i$.
Similarly, $l_0,\dots,l_k$ also decompose $R^\text{sw}$ into $k$ regions $V_1,\dots,V_k$ where $V_i$ is in between $l_{i-1}$ and $l_i$.
See the right figure for an illustration.
For each $i \in [k]$, we report two points $a_i,a_i' \in S \cap U_i$, where $a_i$ is the $L_1$-nearest-neighbor of $o$ among the points in $S \cap U_i$ and $a_i'$ is the $L_1$-nearest-neighbor of $o$ among the points in $S \cap U_i$ whose colors are different from $\text{cl}(a_i)$; we call them the \textit{two $L_1$-NN} of $o$ \textit{with different colors} in $S \cap U_i$.
We shall show later how to build data structures to compute these points efficiently.
Similarly, for each $i \in [k]$, we report two points $b_i,b_i' \in S \cap V_i$ which are the two $L_1$-NN of $o$ with different colors in $S \cap V_i$.
Let $T = \{a_i,a_i',b_i,b_i': i \in [k]\}$.
We then compute $\phi_1 = \mathsf{CP}_\delta(\mathsf{Bich}_o(T))$ by brute-force in $O(k^2)$ time (note that $|T| = 4k$).
By applying Lemma~\ref{lem-angle}, we can prove that $\phi_1$ is a $(1+\varepsilon)$-ACP in $\varPi_1 \Cap R$.
\begin{lemma} \label{lem-correct}
We have $\phi_1 \in \varPi_1 \Cap R$ and $|\phi_1|_\delta \leq (1+\varepsilon) \cdot |\mathsf{CP}_\delta(\varPi_1 \Cap R)|_\delta$.
\end{lemma}
\textit{Proof sketch.}
Suppose $\{a,b\} = \mathsf{CP}_\delta(\varPi_1 \Cap R)$ where $a \in U_i$ and $b \in V_j$.
The key observation here is that we can always find $a^* \in \{a_i,a_i'\}$ and $b^* \in \{b_j,b_j'\}$ such that \textbf{(i)} $\text{cl}(a^*) \neq \text{cl}(b^*)$ and \textbf{(ii)} $L_1(a^*,o) \leq L_1(a,o)$ and $L_1(b^*,o) \leq L_1(b,o)$.
Set $\phi^* = \{a^*,b^*\}$.
By Lemma~\ref{lem-angle}, $|\phi^*|_\delta \leq (1+\varepsilon) \cdot \delta(a,b)$.
Since $\phi_1 = \mathsf{CP}_\delta(\mathsf{Bich}_o(T))$ and $\phi^* \in \mathsf{Bich}_o(T)$, $|\phi_1|_\delta \leq |\phi^*|_\delta \leq (1+\varepsilon) \cdot \delta(a,b)$.
\hfill $\Box$
\smallskip

Next, we discuss how to compute $a_i,a_i'$ and $b_i,b_i'$ for $i \in [k]$ efficiently.
It suffices to consider the points $a_i,a_i'$.
Here we use a special property of the $L_1$-metric.
We observe that, for a point $a \in R^\text{ne}$, the smaller $\mathsf{x}(a)+\mathsf{y}(a)$ is, the smaller $L_1(a,o)$ is.
Note that the value $\mathsf{x}(a)+\mathsf{y}(a)$ is independent of $o$.
As such, if we assign each point $a \in S$ a \textit{weight} equal to $\mathsf{x}(a)+\mathsf{y}(a)$, then $a_i,a_i'$ are just the two lightest points contained in $U_i$ with different colors.
In this sense, computing $a_i,a_i'$ is somehow similar to a range-minimum query with the query range $U_i$, and we solve the problem as follows.
For each $i \in [k]$, we shall build three data structures $\mathbf{C}_i,\mathbf{C}_i^\text{v},\mathbf{C}_i^\text{h}$.
In the query $(R,o)$, if $U_i$ is a 4-gon, then we will use $\mathbf{C}_i$ to compute $a_i,a_i'$.
If $U_i$ is a triangle whose edge coinciding with the boundary of $R$ is vertical (resp., horizontal), we will use $\mathbf{C}_i^\text{v}$ (resp., $\mathbf{C}_i^\text{h}$) to compute $a_i,a_i'$.
To design these data structures, we introduce the following lemma.
\begin{lemma} \label{lem-top2ds}
Let $c \geq 2$ be a constant, $H_1,\dots,H_c$ be halfplanes.
Given a weighted colored dataset $A$ in $\mathbb{R}^2$ of size $m$, one can build an $O(m \log^{c-1} m)$-space data structure on $A$ that can report in $O(\log^{c-1} m)$ time, for any specified $c$ halfplanes $H_1^\dagger,\dots,H_c^\dagger$ where $H_i^\dagger$ is a translate of $H_i$, two points $a,a'$ where $a$ is the lightest point in $A \cap (\bigcap_{i=1}^c H_i^\dagger)$ and $a'$ is the lightest point in $A \cap (\bigcap_{i=1}^c H_i^\dagger)$ whose color is different from $\textnormal{cl}(a)$.
\end{lemma}
\textit{Proof sketch.}
The range-search problem in the lemma is decomposable, and the halfplanes in a query are the translates of $c$ fixed halfplanes.
This allows us to use range-tree-based techniques.
Assume $H_c$ has the equation $x = 0$.
We build a (1D) range tree on $A$ for $x$-coordinates, and recursively build at each node a data structure for $H_1,\dots,H_{c-1}$.
When there are only two halfplanes remaining, we apply fractional cascading to shave off a log-factor in the query time.
\hfill $\Box$
\smallskip

\noindent
As mentioned before, we assign each point $a \in S$ a weight $\mathsf{x}(a)+\mathsf{y}(a)$, then $a_i$ is just the lightest point in $S \cap U_i$ and $a_i'$ is the lightest point in $S \cap U_i$ whose color is different from $\text{cl}(a)$.
This allows us to apply the above lemma.
Note that if $U_i$ is a 4-gon, then it is always the intersection of four halfplanes which are translates of $H_1:x \leq 0$, $H_2:y \leq 0$, $H_3:x \cdot \cos((i-1)\pi/2k) - y \cdot \sin((i-1)\pi/2k) \geq 0$, and $H_4:x \cdot \cos(i\pi/2k) - y \cdot \sin(i\pi/2k) \leq 0$, respectively.
Thus, Lemma~\ref{lem-top2ds} gives us the data structure $\mathbf{C}_i$ with $O(n \log^3 n)$ space and $O(\log^3 n)$ query time.
Similarly, we can build $\mathbf{C}_i^\text{v}$ and $\mathbf{C}_i^\text{h}$ with $O(n \log^2 n)$ space and $O(\log^2 n)$ query time by Lemma~\ref{lem-top2ds}.
The space cost of all these data structures is $O(k n \log^3 n)$.
The total time for computing the points $a_1,a_1',\dots,a_k,a_k'$ is $O(\log^3 n + k \log^2 n)$, because exactly one of $U_1,\dots,U_k$ is a 4-gon and the others are triangles.
The points $b_1,b_1',\dots,b_k,b_k'$ can be computed by building similar data structures.
Including the $O(k^2)$ time for finding $\mathsf{CP}_\delta(\mathsf{Bich}_o(T))$, the time for computing $\phi_1$ is $O(\log^3 n + k \log^2 n + k^2)$, which is also the query time of our anchored CRCP data structure.
Since $k = O(\varepsilon^{-1})$, we conclude the following.
\begin{theorem*} \label{thm-anchored}
Let $\mathcal{R}$ be the collection of rectangles in $\mathbb{R}^2$, and $\delta$ be a metric on $\mathbb{R}^2$ induced by a monotone norm.
For a parameter $\varepsilon>0$, there exists a $(1+\varepsilon)$-approximate anchored $(\mathcal{R},\delta)$-CRCP data structure using $O(\varepsilon^{-1} n \log^3 n)$ space and $O(\log^3 n + \varepsilon^{-1} \log^2 n + \varepsilon^{-2})$ query time.
\end{theorem*}

The result above can be generalized to $\mathbb{R}^d$ for $d \geq 3$.
However, for simplicity of the paper, we only discuss its generalization in $\mathbb{R}^3$, which is already sufficient for our purpose.
\begin{theorem} \label{thm-anchored3d}
Let $\mathcal{B}$ be the collection of boxes in $\mathbb{R}^3$, and $\delta$ be a metric on $\mathbb{R}^3$ induced by a monotone norm.
For a parameter $\varepsilon>0$, there exists a $(1+\varepsilon)$-approximate anchored $(\mathcal{B},\delta)$-CRCP data structure using $O(\varepsilon^{-2} n \log^6 n)$ space and $O(\varepsilon^{-2} \log^6 n + \varepsilon^{-4})$ query time.
\end{theorem}

\section{CRCP data structures in $\mathbb{R}^2$} \label{sec-2d}
Based on the techniques developed in the previous sections, we study the CRCP problem in $\mathbb{R}^2$.
We first consider strip and quadrant queries.
Then, using our strip and quadrant data structures as sub-routines, we obtain our main result, two $(1+\varepsilon)$-approximate CRCP data structures for rectangle queries.
Fix a metric $\delta$ on $\mathbb{R}^2$, and assume $\delta$ is induced by a monotone norm.
The discussion in this section is always under the metric $\delta$.
Let $\varepsilon>0$ be a pre-specified parameter, and $S$ be the given colored dataset in $\mathbb{R}^2$ of size $n$.


\subsection{Strip and quadrant queries} \label{sec-strquad}
Let $\mathcal{P}$ (resp., $\mathcal{Q}$) be the collection of strips (resp., quadrants) in $\mathbb{R}^2$.
To handle these queries, we use the RCP-coreset technique in Section~\ref{sec-coreset} and the following result given in \cite{xue2018new}.
\begin{lemma} \label{lem-xue}
\textnormal{\bf \cite{xue2018new}}
Let $\mathcal{X} = \mathcal{P}$ or $\mathcal{X} = \mathcal{Q}$.
Given a weighted set $\varPi'$ of $m$ point pairs in $\mathbb{R}^2$, one can build an $O(m)$-space data structure which can return in $O(\log m)$ time, for a specified query range $X \in \mathcal{X}$, the lightest pair in $\varPi' \Cap X$.
\end{lemma}

We first consider the strip queries.
Clearly, it suffices to handle the queries in the sub-collection $\mathcal{P}^\text{v} \subseteq \mathcal{P}$ consisting of vertical strips.
Since $\mathcal{P}^\text{v}$ is well-behaved on any set of point pairs in $\mathbb{R}^2$ (see Appendix~\ref{appx-well}) and thus on $\mathsf{Bich}(S)$, there exists an $\varepsilon$-coreset $\varPi'$ of $\mathsf{Bich}(S)$ for $\mathcal{P}^\text{v}$-queries of size $O(\varepsilon^{-1} n \log^2 n)$ by Theorem~\ref{thm-coreset}.
We then build a data structure of Lemma~\ref{lem-xue} on $\varPi'$ for $\mathcal{X} = \mathcal{P}$, by assigning each $\phi \in \varPi'$ a weight $|\phi|_\delta$.
This gives us the approximate CRCP data structure, because for any $P \in \mathcal{P}^\text{v}$, the lightest pair in $\varPi' \Cap P$ is $\mathsf{CP}_\delta(\varPi' \Cap P)$, which is a $(1+\varepsilon)$-ACP in $\mathsf{Bich}(S) \Cap P$.
\begin{theorem*} \label{thm-strip}
Let $\mathcal{P}$ be the collection of strips in $\mathbb{R}^2$, and $\delta$ be a metric on $\mathbb{R}^2$ induced by a monotone norm.
For a parameter $\varepsilon>0$, there exists a $(1+\varepsilon)$-approximate $(\mathcal{P},\delta)$-CRCP data structure using $O(\varepsilon^{-1} n \log^2 n)$ space and $O(\log n + \log (1/\varepsilon))$ query time.
\end{theorem*}

Next, we consider the quadrant queries.
It suffices to handle the queries in the sub-collection $\mathcal{Q}^\text{ne} \subseteq \mathcal{Q}$ consisting of northeast quadrants, i.e., quadrants of the form $[x^-,+\infty) \times [y^-,+\infty)$.
Unfortunately, $\mathcal{Q}^\text{ne}$ may be not well-behaved on $\mathsf{Bich}(S)$.
To resolve this issue, we write $\mathsf{Bich}(S) = \varPi_1 \cup \varPi_2$ where $\varPi_1$ (resp., $\varPi_2$) consists of the NW-SE (resp., NE-SW) pairs in $\mathsf{Bich}(S)$.
Since $\mathcal{Q}^\text{ne}$ is well-behaved on any set of NW-SE pairs in $\mathbb{R}^2$ (see Appendix~\ref{appx-well}) and thus on $\varPi_1$, there exists an $\varepsilon$-coreset $\varPi_1'$ of $\varPi_1$ for $\mathcal{Q}^\text{ne}$-queries of size $O(\varepsilon^{-1} n \log^2 n)$ by Theorem~\ref{thm-coreset}.
Define $\varPi' = \varPi_1' \cup \varPi_2'$ where $\varPi_2' = \{\mathsf{CP}_\delta(\varPi_2 \Cap Q): Q \in \mathcal{Q}^\text{ne}\}$.
Then $\varPi'$ is an $\varepsilon$-coreset of $\mathsf{Bich}(S)$, because $\varPi_1'$ (resp., $\varPi_2'$) is an $\varepsilon$-coreset of $\varPi_1$ (resp., $\varPi_2$).
We then build a data structure of Lemma~\ref{lem-xue} on $\varPi'$ for $\mathcal{X} = \mathcal{Q}$ by assigning each $\phi \in \varPi'$ a weight $|\phi|_\delta$, which gives us the approximate CRCP data structure.
The performance of the data structure depends on $|\varPi'|$, which is analyzed in the following lemma.
\begin{lemma} \label{lem-candOn}
We have $|\varPi_2'| \leq n$ and thus $|\varPi'| = O(\varepsilon^{-1} n \log^2 n)$.
\end{lemma}
\textit{Proof sketch.}
Let $\phi = \{a,b\} \in \varPi_2'$ where $a$ is the southwest point of $\phi$.
The observation is that $b$ must be the $\delta$-nearest-neighbor of $a$ among all the points to the northeast of $a$ whose colors are different from $\text{cl}(a)$.
Thus, if we charge each pair in $\varPi_2'$ to its southwest point, every point in $S$ is charged at most once, which implies $|\varPi_2'| \leq n$.
\hfill $\Box$

\begin{theorem*} \label{thm-quadrant}
Let $\mathcal{Q}$ be the collection of quadrants in $\mathbb{R}^2$, and $\delta$ be a metric on $\mathbb{R}^2$ induced by a monotone norm.
For a parameter $\varepsilon>0$, there exists a $(1+\varepsilon)$-approximate $(\mathcal{Q},\delta)$-CRCP data structure using $O(\varepsilon^{-1} n \log^2 n)$ space and $O(\log n + \log (1/\varepsilon))$ query time.
\end{theorem*}


\subsection{Rectangle query} \label{sec-rectangle}
\begin{wrapfigure}{r}{4.2cm}
    \vspace{-0.6cm}
	\begin{center}
		\includegraphics[height=3.5cm]{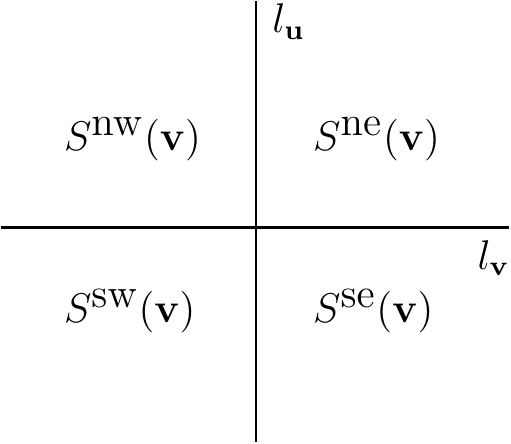}
	\end{center}
	\vspace{-0.6cm}
	\label{fig-4subset}
\end{wrapfigure}
Let $\mathcal{R}$ be the collection of (axis-parallel) rectangles in $\mathbb{R}^2$.
In this section, we give our main result, i.e., two $(1+\varepsilon)$-approximate $(\mathcal{R},\delta)$-CRCP data structures, based on our strip/quadrant CRCP data structures (Section~\ref{sec-strquad}) and anchored CRCP data structure (Section~\ref{sec-anchored}).

Our first data structure consists of three parts.
The first part is a standard 2D range tree $\mathbf{T}$ built on $S$ \cite{bkos_2000}.
The main tree (or primary tree) $\mathbf{T}_*$ of $\mathbf{T}$ is a (1D) range tree built on the $x$-coordinates of the points in $S$.
The leaves of $\mathbf{T}_*$ one-to-one correspond to the points in $S$.
Each node $\mathbf{u} \in \mathbf{T}_*$ corresponds to a \textit{cannonical subset} $S(\mathbf{u}) \subseteq S$ consisting of all points stored at the leaves of the subtree rooted at $\mathbf{u}$.
At each node $\mathbf{u} \in \mathbf{T}_*$, there is an associated secondary tree $\mathbf{T}_\textbf{u}$, which is a (1D) range tree built on the $y$-coordinates of the points in $S(\mathbf{u})$.
With an abuse of notation, for a node $\mathbf{v} \in \mathbf{T}_\textbf{u}$, we also use $S(\mathbf{v})$ to denote the canonical subset of $\mathbf{v}$, which is a subset of $S(\mathbf{u})$.
For each internal primary node $\mathbf{u} \in \mathbf{T}_*$, we fix a vertical line $l_\mathbf{u}$ such that the points in the left (resp., right) subtree of $\mathbf{u}$ are to the left (resp., right) of $l_\mathbf{u}$.
Similarly, for each internal secondary node $\mathbf{v}$, we fix a horizontal line $l_\mathbf{v}$ such that the points in the left (resp., right) subtree of $\mathbf{v}$ are below (resp., above) $l_\mathbf{u}$.
Let $\mathbf{v} \in \mathbf{T}_\textbf{u}$ be an internal secondary node.
We associate to $\mathbf{v}$ the two lines $l_\mathbf{u}$ and $l_\mathbf{v}$, which separate $S(\mathbf{v})$ into four subsets $S^\text{ne}(\mathbf{v}),S^\text{nw}(\mathbf{v}),S^\text{sw}(\mathbf{v}),S^\text{se}(\mathbf{v})$; see the right figure.
We build $(1+\varepsilon)$-approximate $(\mathcal{Q},\delta)$-CRCP data structures (Theorem~\ref{thm-quadrant}) on $S^\text{ne}(\mathbf{v}),S^\text{nw}(\mathbf{v}),S^\text{sw}(\mathbf{v}),S^\text{se}(\mathbf{v})$ respectively, and denote them by $\mathbf{A}^\text{ne}(\mathbf{v}),\mathbf{A}^\text{nw}(\mathbf{v}),\mathbf{A}^\text{sw}(\mathbf{v}),\mathbf{A}^\text{se}(\mathbf{v})$.
We store these four data structures at $\mathbf{v}$.
It is easy to see that the space cost of $\mathbf{T}$ is $O(\varepsilon^{-1} n \log^4 n)$, since the space of each $\mathbf{T}_\mathbf{u}$ is $O(\varepsilon^{-1} |S(\mathbf{u})| \log^3 |S(\mathbf{u})|)$.

The second part of our data structure are two 1D range trees $\mathbf{T}'$ and $\mathbf{T}''$ built on $S$, where $\mathbf{T}'$ (resp., $\mathbf{T}''$) is built on the $x$-coordinates (resp., $y$-coordinates).
Again, for a node $\mathbf{u}$ of $\mathbf{T}'$ or $\mathbf{T}''$, we use the notation $S(\mathbf{u})$ to denote the canonical subset of $\mathbf{u}$.
At each node $\mathbf{u}' \in \mathbf{T}'$ (resp., $\mathbf{u}'' \in \mathbf{T}''$), we store a $(1+\varepsilon)$-approximate $(\mathcal{P},\delta)$-CRCP data structure (Theorem~\ref{thm-strip}) built on $S(\mathbf{u}')$ (resp., $S(\mathbf{u}'')$), denoted by $\mathbf{B}(\mathbf{u}')$ (resp., $\mathbf{B}(\mathbf{u}'')$).
The space cost of $\mathbf{T}'$ and $\mathbf{T}''$ is clearly $O(\varepsilon^{-1} n \log^3 n)$.

The third part of our data structure is a $(1+\varepsilon)$-approximate anchored $(\mathcal{R},\delta)$-CRCP data structure $\mathbf{C}$ built on $S$ (Theorem~\ref{thm-anchored}), which uses $O(\varepsilon^{-1} n \log^3 n)$ space.

\begin{wrapfigure}{r}{4.2cm}
    \vspace{-0.6cm}
	\begin{center}
		\includegraphics[height=3.5cm]{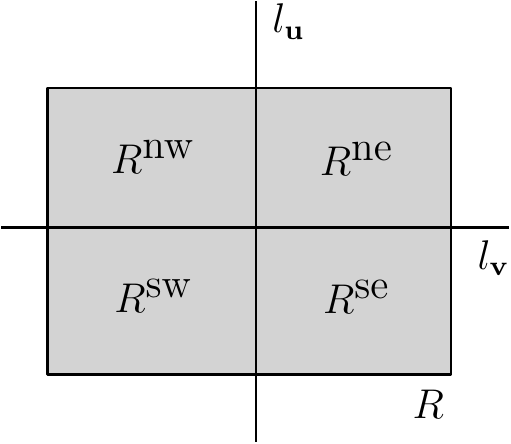}
	\end{center}
	\vspace{-0.6cm}
	\label{fig-4rect}
\end{wrapfigure}
The overall space cost of the entire data structure is $O(\varepsilon^{-1} n \log^4 n)$.
Next, we show how to answer a query $R = [x^-,x^+] \times [y^-,y^+] \in \mathcal{R}$ using the above data structure.
We first consider the 2D range tree $\mathbf{T}$.
We find in the primary tree $\mathbf{T}_*$ the \textit{splitting node} $\mathbf{u} \in \mathbf{T}_*$ corresponding to the range $[x^-,x^+]$, which is by definition the LCA of all the leaves whose corresponding points have $x$-coordinates in $[x^-,x^+]$.
Then we find in the secondary tree $\mathbf{T}_\mathbf{u}$ the splitting node $\mathbf{v} \in \mathbf{T}_\mathbf{u}$ corresponding to the range $[y^-,y^+]$.
By the property of splitting node, we have $S \cap R = S(\mathbf{v}) \cap R$, and the lines $l_\mathbf{u}$ and $l_\mathbf{v}$ both intersect $R$.
Thus, $l_\mathbf{u}$ and $l_\mathbf{v}$ decompose $R$ into four small rectangles $R^\text{ne},R^\text{nw},R^\text{sw},R^\text{se}$; see the right figure.
Define $Q^\text{ne}$ as the quadrant obtained by removing the two sides of $R^\text{ne}$ that coincide with $l_\mathbf{u}$ and $l_\mathbf{v}$, i.e., $Q^\text{ne} = [x^-,\infty) \times [y^-,\infty)$, etc.
Then $S^\text{ne}(\mathbf{v}) \cap R = S^\text{ne}(\mathbf{v}) \cap Q^\text{ne}$.
Similarly, we define $Q^\text{nw},Q^\text{sw},Q^\text{se}$.
Next, we consider the two 1D range trees $\mathbf{T}'$ and $\mathbf{T}''$.
In $\mathbf{T}'$, we find the canonical nodes corresponding to the range $[x^-,x^+]$, say $\mathbf{u}'_1,\dots,\mathbf{u}'_{t'} \in \mathbf{T}'$ where $t' = O(\log n)$.
Assume $\mathbf{u}'_1,\dots,\mathbf{u}'_{t'}$ are sorted from left to right in $\mathbf{T}'$, and thus the points in $S(\mathbf{u}'_i)$ have smaller $x$-coordinates than those in $S(\mathbf{u}'_j)$ for $i<j$.
We find $t'-1$ real numbers $x_1,\dots,x_{t'-1}$ which ``separate'' $S(\mathbf{u}'_1),\dots,S(\mathbf{u}'_{t'})$, i.e., $x_i$ is greater than (resp., smaller than) the $x$-coordinates of the points in $S(\mathbf{u}'_i)$ (resp., $S(\mathbf{u}'_{i+1})$).
Similarly, we find the canonical nodes in $\mathbf{T}''$ corresponding to range $[y^-,y^+]$, say $\mathbf{u}''_1,\dots,\mathbf{u}''_{t''} \in \mathbf{T}''$ where $t'' = O(\log n)$, and $t''-1$ real numbers $y_1,\dots,y_{t''-1}$ which separate $S(\mathbf{u}''_1),\dots,S(\mathbf{u}''_{t''})$.
Set $O_1 = \{o'_1,\dots,o'_{t'-1},o''_1,\dots,o''_{t''-1}\}$, where $o'_i$ is the point on $l_\mathbf{v}$ with $x$-coordinate $x_i$ and $o''_i$ is the point on $l_\mathbf{u}$ with $x$-coordinate $y_i$.
We do the following queries. \\
$\bullet$ We query $\mathbf{A}^\text{ne}(\mathbf{v}),\mathbf{A}^\text{nw}(\mathbf{v}),\mathbf{A}^\text{sw}(\mathbf{v}),\mathbf{A}^\text{se}(\mathbf{v})$ with the quadrants $Q^\text{ne},Q^\text{nw},Q^\text{sw},Q^\text{se}$, respectively. \\
$\bullet$ We query $\mathbf{B}(\mathbf{u}'_i)$ with the strip $\mathbb{R} \times [y^-,y^+]$ for all $i \in [t']$. \\
$\bullet$ We query $\mathbf{B}(\mathbf{u}''_i)$ with the strip $[x^-,x^+] \times \mathbb{R}$ for all $i \in [t'']$. \\
$\bullet$ We query $\mathbf{C}$ with $(R,o)$ for all $o \in O_1$ (in total $t'+t''-1$ queries). \\
Denote by $\varPhi_1$ the set of all pairs returned by the above queries.
We then return $\mathsf{CP}_\delta(\varPhi_1)$ as our final answer for the query $R$.
It is easy to see that the entire query time is $O(\log^4n + \varepsilon^{-1} \log^3n + \varepsilon^{-2} \log n)$.
Indeed, the most time-consuming part is the query on $\mathbf{C}$; we have $t'+t''-1 = O(\log n)$ such queries.
It suffices to verify the correctness of the answer.
\begin{lemma} \label{lem-rect1correct}
The pair $\mathsf{CP}_\delta(\varPhi_1)$ is a $(1+\varepsilon)$-ACP in $\mathsf{Bich}(S) \Cap R$ under the metric $\delta$.
\end{lemma}
\textit{Proof.}
It suffices to show some $\phi \in \varPhi_1$ is a $(1+\varepsilon)$-ACP in $\mathsf{Bich}(S) \Cap R$.
Let $\phi^* = \mathsf{CP}_\delta(\mathsf{Bich}(S) \Cap R)$.
If the two points of $\phi^*$ are in one of $S^\text{ne}(\mathbf{v}),S^\text{nw}(\mathbf{v}),S^\text{sw}(\mathbf{v}),S^\text{se}(\mathbf{v})$, the answer returned by the corresponding $\mathbf{A}^\bullet(\mathbf{v})$ is a $(1+\varepsilon)$-ACP.
If $\phi^* \subseteq S(\mathbf{u}'_i)$ for some $i \in [t']$, the answer returned by $\mathbf{B}(\mathbf{u}'_i)$ is a $(1+\varepsilon)$-ACP.
If $\phi^* \subseteq S(\mathbf{u}''_i)$ for some $i \in [t'']$, the answer returned by $\mathbf{B}(\mathbf{u}''_i)$ is a $(1+\varepsilon)$-ACP.
If all of the above are not true, it is easy to see that $\phi^*$ is $o$-anchored for some $o \in O_1$.
Thus, the query $(R,o)$ on $\mathbf{C}$ returns a $(1+\varepsilon)$-ACP.
\hfill $\Box$

\begin{theorem*} \label{thm-rect1}
Let $\mathcal{R}$ be the collection of rectangles in $\mathbb{R}^2$, and $\delta$ be a metric on $\mathbb{R}^2$ induced by a monotone norm.
For a parameter $\varepsilon>0$, there exists a $(1+\varepsilon)$-approximate $(\mathcal{R},\delta)$-CRCP data structure using $O(\varepsilon^{-1} n \log^4 n)$ space and $O(\log^4n + \varepsilon^{-1} \log^3n + \varepsilon^{-2}\log n)$ query time.
\end{theorem*}

Our second data structure is a variant of the first one, and is obtained by simply removing the 2D range tree $\mathbf{T}$ from the first data structure (i.e., it consists of the two 1D range trees $\mathbf{T}',\mathbf{T}''$ and the anchored data structure $\mathbf{C}$).
Thus, its space cost is $O(\varepsilon^{-1}n \log^3 n)$.
To answer a query $R = [x^-,x^+] \times [y^-,y^+] \in \mathcal{R}$, we do the same thing on $\mathbf{T}'$ and $\mathbf{T}''$ as before, namely, find the canonical nodes $\mathbf{u}'_1,\dots,\mathbf{u}'_{t'} \in \mathbf{T}'$, $\mathbf{u}''_1,\dots,\mathbf{u}''_{t''} \in \mathbf{T}''$ and the numbers $x_1,\dots,x_{t'-1}$, $y_1,\dots,y_{t''-1}$.
Set $O_2 = \{o_{i,j}: i \in [t'-1] \text{ and } j \in [t''-1]\}$ where $o_{i,j} = (x_i,y_j)$.
We do the following queries. \\
$\bullet$ We query $\mathbf{B}(\mathbf{u}'_i)$ with the strip $\mathbb{R} \times [y^-,y^+]$ for all $i \in [t']$. \\
$\bullet$ We query $\mathbf{B}(\mathbf{u}''_i)$ with the strip $[x^-,x^+] \times \mathbb{R}$ for all $i \in [t'']$. \\
$\bullet$ We query $\mathbf{C}$ with $(R,o)$ for all $o \in O_2$ (in total $O(t't'')$ queries). \\
Denote by $\varPhi_2$ the set of all pairs returned by the above queries.
We then return $\mathsf{CP}_\delta(\varPhi_2)$ as our final answer .
The entire query time is $O(\log^5n + \varepsilon^{-1} \log^4n + \varepsilon^{-2} \log^2 n)$, since we do $O(t't'') = O(\log^2 n)$ queries on $\mathbf{C}$ at this time.
It suffices to verify the correctness of the answer.
\begin{lemma} \label{lem-rect2correct}
The pair $\mathsf{CP}_\delta(\varPhi_2)$ is a $(1+\varepsilon)$-ACP in $\mathsf{Bich}(S) \Cap R$ under the metric $\delta$.
\end{lemma}

\begin{theorem*} \label{thm-rect2}
Let $\mathcal{R}$ be the collection of rectangles in $\mathbb{R}^2$, and $\delta$ be a metric on $\mathbb{R}^2$ induced by a monotone norm.
For a parameter $\varepsilon>0$, there exists a $(1+\varepsilon)$-approximate $(\mathcal{R},\delta)$-CRCP data structure using $O(\varepsilon^{-1} n \log^3 n)$ space and $O(\log^5n + \varepsilon^{-1} \log^4n + \varepsilon^{-2}\log^2 n)$ query time.
\end{theorem*}

\noindent
\textbf{Remark.}
The solution in \cite{xue2018new} for the rectangle RCP problem also uses range trees to reduce a rectangle query into several strip/quadrant queries.
Unfortunately, that reduction does not work for the CRCP problem.
Indeed, it relies on some geometric observations similar to those in the standard divide-and-conquer algorithm for the classical closest-pair problem, e.g., for a set $A$ of points with closest-pair distance $\sigma$, any rectangle of size $O(\sigma) \times O(\sigma)$ can only contain $O(1)$ points in $A$.
These observations are inapplicable to the colored version.
As such, we use different reductions in this section, which take advantage of our anchored CRCP data structure.

\section{CRCP problem in higher dimensions} \label{sec-higher}
By applying our techniques in the previous sections, we also achieve some results for the CRCP problem in higher dimensions, which is presented in this section.
Specifically, we consider the CRCP problem with slab and 2-box queries in $\mathbb{R}^d$ (for a constant $d$), and with dominance query in $\mathbb{R}^3$.
Fix a metric on $\mathbb{R}^d$, and assume $\delta$ is induced by a monotone norm.
The discussion in this section is always under the metric $\delta$.
Let $\varepsilon>0$ be a pre-specified parameter, and $S$ be the given colored dataset in $\mathbb{R}^d$ of size $n$.

\subsection{Slab and 2-box queries in $\mathbb{R}^d$}
A \textit{slab} in $\mathbb{R}^d$ is an orthogonal range bounded on two sides in exactly one dimension (and unbounded in other dimensions); a slab in $\mathbb{R}^2$ is just a strip.
A \textit{2-box} in $\mathbb{R}^d$ is an orthogonal range bounded on one side in exactly two dimensions; a 2-box in $\mathbb{R}^2$ is just a quadrant.
Let $\mathcal{L}$ (resp., $\mathcal{B}_2$) be the collection of slabs (resp., 2-boxes) in $\mathbb{R}^d$.
Since the slab and 2-box queries are generalizations of the strip and quadrant queries, our approach here is the same as in Section~\ref{sec-strquad}.
We first give a generalization of Lemma~\ref{lem-xue}.
\begin{lemma} \label{lem-xue'}
Let $\mathcal{X} = \mathcal{L}$ or $\mathcal{X} = \mathcal{B}_2$.
Given a weighted set $\varPi'$ of $m$ point pairs in $\mathbb{R}^d$, one can build an $O(m)$-space data structure which can return in $O(\log m)$ time, for a specified query range $X \in \mathcal{X}$, the lightest pair in $\varPi' \Cap X$.
\end{lemma}

We first consider the slab queries.
Clearly, it suffices to handle the queries in the sub-collection $\mathcal{L}' \subseteq \mathcal{L}$ consisting of slabs of the form $[x^-,x^+] \times \mathbb{R}^{d-1}$ (i.e., bounded in the first dimension).
Since $\mathcal{L}'$ is well-behaved on any set of point pairs in $\mathbb{R}^d$ (see Appendix~\ref{appx-well}) and thus on $\mathsf{Bich}(S)$, there exists an $\varepsilon$-coreset of $\mathsf{Bich}(S)$ for $\mathcal{L}'$-queries of size $O(\varepsilon^{-1}n\log^dn)$ by Theorem~\ref{thm-coreset}.
We then build a data structure of Lemma~\ref{lem-xue'} on $\varPi'$ for $\mathcal{X} = \mathcal{L}$, which gives us the approximate CRCP data structure.
\begin{theorem*} \label{thm-slab}
Let $\mathcal{L}$ be the collection of slabs in $\mathbb{R}^d$, and $\delta$ be a metric on $\mathbb{R}^d$ induced by a monotone norm.
For a parameter $\varepsilon>0$, there exists a $(1+\varepsilon)$-approximate $(\mathcal{L},\delta)$-CRCP data structure using $O(\varepsilon^{-1} n \log^d n)$ space and $O(\log n + \log(1/\varepsilon))$ query time.
\end{theorem*}

Next, we consider the 2-box queries.
It suffices to handle the queries in the sub-collection $\mathcal{B}_2' \subseteq \mathcal{B}$ consisting of 2-boxes of the form $[x^-,\infty) \times [y^-,\infty) \times \mathbb{R}^{d-2}$.
Define a map $\pi: \mathbb{R}^d \rightarrow \mathbb{R}^2$ as $(x_1,\dots,x_d) \mapsto (x_1,x_2)$.
We write $\mathsf{Bich}(S) = \varPi_1 \cup \varPi_2$ where $\varPi_1 = \{\phi \in \mathsf{Bich}(S): \pi(\phi) \text{ is a NW-SE pair}\}$ and $\varPi_2 = \{\phi \in \mathsf{Bich}(S): \pi(\phi) \text{ is a NE-SW pair}\}$.
Now $\mathcal{B}_2'$ is well-behaved on $\varPi_1$ (see Appendix~\ref{appx-well}).
Thus, by Theorem~\ref{thm-coreset}, there exists an $\varepsilon$-coreset $\varPi_1'$ of $\varPi_1$ for $\mathcal{B}_2'$-queries of size $O(\varepsilon^{-1}n\log^dn)$.
Define $\varPi' = \varPi_1' \cup \varPi_2'$ where $\varPi_2' = \{\mathsf{CP}_\delta(\varPi_2 \Cap B): B \in \mathcal{B}_2'\}$.
Then $\varPi'$ is an $\varepsilon$-coreset of $\mathsf{Bich}(S)$, because $\varPi_1'$ (resp., $\varPi_2'$) is an $\varepsilon$-coreset of $\varPi_1$ (resp., $\varPi_2$).
We then build a data structure of Lemma~\ref{lem-xue'} on $\varPi'$ for $\mathcal{X} = \mathcal{B}_2$, which gives us the approximate CRCP data structure.
The performance of the data structure depends on $|\varPi'|$, which is analyzed in the following lemma.
\begin{lemma} \label{lem-candOnRd}
We have $|\varPi_2'| \leq n$ and thus $|\varPi'| = O(\varepsilon^{-1}n\log^dn)$.
\end{lemma}
\begin{theorem*} \label{thm-2box}
Let $\mathcal{B}_2$ be the collection of 2-boxes in $\mathbb{R}^d$, and $\delta$ be a metric on $\mathbb{R}^d$ induced by a monotone norm.
For a parameter $\varepsilon>0$, there exists a $(1+\varepsilon)$-approximate $(\mathcal{B}_2,\delta)$-CRCP data structure using $O(\varepsilon^{-1} n \log^d n)$ space and $O(\log n + \log(1/\varepsilon))$ query time.
\end{theorem*}

\subsection{Dominance query in $\mathbb{R}^3$}
Suppose $d=3$ in this section.
A \textit{dominance region} in $\mathbb{R}^3$ is a range of the form $[x^-,\infty) \times [y^-,\infty) \times [z^-,\infty)$.
Let $\mathcal{D}$ be the collection of dominance regions in $\mathbb{R}^3$.

Our idea for solving the $(\mathcal{D},\delta)$-CRCP problem is similar to our second data structure for rectangle queries (Section~\ref{sec-rectangle}).
We build three (1D) range trees $\mathbf{T}_x,\mathbf{T}_y,\mathbf{T}_z$ on $S$, where $\mathbf{T}_x$ is built on the $x$-coordinates, $\mathbf{T}_y$ is built on the $y$-coordinates, and $\mathbf{T}_z$ is built on the $z$-coordinates.
At each node $\mathbf{u} \in \mathbf{T}_x$, we store a $(1+\varepsilon)$-approximate $(\mathcal{B}_2,\delta)$-CRCP data structure (Theorem~\ref{thm-2box}) built on the canonical subset $S(\mathbf{u})$ of $\mathbf{u}$, which uses $O(\varepsilon^{-1} |S(\mathbf{u})| \log^3 |S(\mathbf{u})|)$ space.
We do the same thing for each node in $\mathbf{T}_y$ and $\mathbf{T}_z$.
The overall space cost of $\mathbf{T}_x,\mathbf{T}_y,\mathbf{T}_z$ is then $O(\varepsilon^{-1} n \log^4 n)$.
Besides the three range trees, we also build a $(1+\varepsilon)$-approximate anchored $(\mathcal{B},\delta)$-CRCP data structure $\mathbf{C}$ (Theorem~\ref{thm-anchored3d}) on $S$, which uses $O(\varepsilon^{-2} n \log^6 n)$ space.

We now show how to answer a query $D = [x^-,\infty) \times [y^-,\infty) \times [z^-,\infty) \in \mathcal{D}$ using the above data structure.
In $\mathbf{T}_x$, we find the $t = O(\log n)$ canonical nodes $\mathbf{u}_1,\dots,\mathbf{u}_t$ corresponding to the range $[x^-,\infty)$.
Assume $\mathbf{u}_1,\dots,\mathbf{u}_t$ are sorted from left to right in $\mathbf{T}_x$, and thus the points in $S(\mathbf{u}_i)$ have smaller $x$-coordinates than those in $S(\mathbf{u}_j)$ for $i<j$.
We find $t-1$ real numbers $x_1,\dots,x_{t-1}$ which ``separate'' $S(\mathbf{u}_1),\dots,S(\mathbf{u}_t)$, i.e., $x_i$ is greater than the $x$-coordinates of the points in $S(\mathbf{u}_i)$ and smaller than the $x$-coordinates of the points in $S(\mathbf{u}_{i+1})$.
Similarly, we find the $t' = O(\log n)$ canonical nodes $\mathbf{v}_1,\dots,\mathbf{v}_{t'} \in \mathbf{T}_y$ corresponding to the range $[y^-,\infty)$, and the $t'' = O(\log n)$ canonical nodes $\mathbf{w}_1,\dots,\mathbf{w}_{t''} \in \mathbf{T}_z$ corresponding to the range $[z^-,\infty)$.
Also, we find real numbers $y_1,\dots,y_{t'-1}$ and $z_1,\dots,z_{t''-1}$ which separate $S(\mathbf{v}_1),\dots,S(\mathbf{v}_{t'})$ and $S(\mathbf{w}_1),\dots,S(\mathbf{w}_{t''})$, respectively.
Set $O = \{o_{i,j,k}: i \in [t-1], j \in [t'-1], k \in [t''-1]\}$ where $o_{i,j,k} = (x_i,y_j,z_k) \in \mathbb{R}^3$.
We do the following queries. \\
$\bullet$ We query the data structure stored at $\mathbf{u}_i$ with the 2-box $\mathbb{R} \times [y^-,\infty) \times [z^-,\infty)$, for all $i \in [t]$. \\
$\bullet$ We query the data structure stored at $\mathbf{v}_i$ with the 2-box $[x^-,\infty) \times \mathbb{R} \times [z^-,\infty)$, for all $i \in [t']$. \\
$\bullet$ We query the data structure stored at $\mathbf{w}_i$ with the 2-box $[x^-,\infty) \times [y^-,\infty) \times \mathbb{R}$, for all $i \in [t'']$. \\
$\bullet$ We query the anchored CRCP data structure $\mathbf{C}$ with $(B,o)$ for all $o \in O$. \\
Denote by $\varPhi$ the set of all pairs returned by the above queries.
We then return $\mathsf{CP}_\delta(\varPhi)$ as our final answer for the query $B$.
It is easy to see that the entire query time is $O(\varepsilon^{-2} \log^9 n + \varepsilon^{-4} \log^3 n)$.
Indeed, the most time-consuming queries part is the queries on $\mathbf{C}$; we have $|O| = O(\log^3 n)$ such queries.
It suffices to verify the correctness of the answer.
\begin{lemma} \label{lem-domcorrect}
The pair $\mathsf{CP}_\delta(\varPhi)$ is a $(1+\varepsilon)$-ACP in $\mathsf{Bich}(S) \Cap B$ under the metric $\delta$.
\end{lemma}

\begin{theorem*} \label{thm-3dom}
Let $\mathcal{D}$ be the collection of dominance regions in $\mathbb{R}^3$, and $\delta$ be a metric on $\mathbb{R}^3$ induced by a monotone norm.
For a parameter $\varepsilon>0$, there exists a $(1+\varepsilon)$-approximate $(\mathcal{D},\delta)$-CRCP data structure using $O(\varepsilon^{-2} n \log^6 n)$ space and $O(\varepsilon^{-2} \log^9 n + \varepsilon^{-4} \log^3 n)$ query time.
\end{theorem*}

\section{Exact solutions might be difficult} \label{sec-difficult}
In this section, we give some evidence showing that designing exact CRCP data structures might be difficult.
Also, we will see that the previous techniques for solving the uncolored RCP problem are inapplicable to the colored version even under the Euclidean metric.

Let $S$ be a given (uncolored) dataset in $\mathbb{R}^2$, and $\mathcal{P}$ (resp., $\mathcal{Q}$) be the collection of strips (resp., quadrants) in $\mathbb{R}^2$.
A pair of points in $S$ is a \textit{candidate pair} if it is the closest pair of points inside some query range.
It was known that the number of the candidate pairs is $O(n \log n)$ for strip queries and $O(n)$ for quadrant queries \cite{gupta2014data,sharathkumar2007range}.
All previous solutions for the strip and quadrant RCP problems are based on these near-linear upper bounds for the number of the candidate pairs; they store all candidate pairs and search for the answer among these pairs \cite{gupta2014data,sharathkumar2007range,xue2018new}.
Furthermore, the solutions for rectangle RCP queries rely on the data structures for strip and quadrant queries.

The concept of candidate pairs can be directly generalized to the CRCP problem (namely, a pair of points in a colored dataset is a candidate pair if it is the closest bichromatic pair of points inside some query range).
Unfortunately, in the CRCP problem, the number of the candidate pairs for strip or quadrant queries is quadratic in the worst-case even in a 2-colored dataset.
\begin{theorem} \label{thm-hard}
There exists a 2-colored dataset $S_1$ \textnormal{(}resp., $S_2$\textnormal{)} in $\mathbb{R}^2$ of size $n$ such that the number of the candidate pairs in $S_1$ \textnormal{(}resp., $S_2$\textnormal{)} for strip \textnormal{(}resp., quadrant\textnormal{)} queries is $\Omega(n^2)$.
\end{theorem}

\noindent
The strip and quadrant queries are already the most ``fundamental'' orthogonal queries.
For these query types, it seems not much one can do beyond storing the candidate pairs (if we want an exact data structure), due to the non-decomposibility of the RCP problem.
As such, the quadratic lower bound given in the above lemma makes it quite difficult (though not impossible) to design efficient exact CRCP data structures, especially under a general metric.






\newpage
\appendix

\noindent
{\LARGE \textbf{Appendix}}

\section{Missing proofs} \label{appx-proof}

\subsection{Proof of Lemma~\ref{lem-mono}}
By definition, we have $\delta(a,b) = \lVert a-b \rVert$.
Suppose $a-b = (x_1,\dots,x_d)$.
Without loss of generality, assume $|x_1| \geq |x_i|$ for all $i \in [d]$.
Then we have $|x_1| \geq (1/\sqrt{d}) \cdot L_2(a,b)$.
Furthermore, $\lVert a-b \rVert \geq \lVert (x_1,0,\dots,0) \rVert = |x_1| \cdot \lVert e_1 \rVert$ by the monotonicity of $\lVert \cdot \rVert$.
Therefore, $\lVert a-b \rVert \geq (1/\sqrt{d}) \cdot L_2(a,b) \cdot \lVert e_1 \rVert \geq (1/\sqrt{d}) \cdot L_2(a,b) \cdot \min_{i \in [d]} \lVert e_i \rVert$.
On the other hand, $\lVert a-b \rVert \leq \sum_{i=1}^d |x_i| \cdot \lVert e_i \rVert \leq |x_1| \sum_{i=1}^d \lVert e_i \rVert$ by the triangle inequality.
Since $|x_1| \leq L_2(a,b)$, we have $|x_1| \sum_{i=1}^d \lVert e_i \rVert \leq d \cdot L_2(a,b) \cdot \max_{i \in [d]} \lVert e_i \rVert$.
Thus, $\lVert a-b \rVert \leq d \cdot L_2(a,b) \cdot \max_{i \in [d]} \lVert e_i \rVert$, as desired.


\subsection{Proof of Lemma~\ref{lem-minimal}}
Let $\mathcal{X}_\varPi = \{X_\phi: \phi \in \varPi\}$.
Since $\mathcal{X}_\varPi$ is finite, there is a minimal element $X^* \in \mathcal{X}_\varPi$.
Note that every $X \in \mathcal{X}_\varPi$ satisfies $\varPi \Cap X \neq \emptyset$ (due to the fact $\phi \subseteq X_\phi$).
Thus, $\varPi \Cap X^* \neq \emptyset$.
We now verify the minimality of $X^*$.
Let $X \in \mathcal{X}$ such that $X \subsetneq X^*$.
We need to show that $\varPi \Cap X = \emptyset$.
Assume $\phi \subseteq X$ for some $\phi \in \varPi$.
Then $X_\phi \subseteq X \subsetneq X^*$ by the definition of $X_\phi$, which contradicts the fact that $X^*$ is a minimal element in $\mathcal{X}_\varPi$.
Therefore, $\varPi \Cap X = \emptyset$, completing the proof.

\subsection{Proof of Lemma~\ref{lem-corecorrect}}
Let $X \in \mathcal{X}$ be an arbitrary range such that $\varPi \Cap X \neq \emptyset$.
We need to verify $\varPi' \Cap X \neq \emptyset$ and the inequality that $|\mathsf{CP}_\delta(\varPi' \Cap X)|_\delta \leq (1+\varepsilon) \cdot |\mathsf{CP}_\delta(\varPi \Cap X)|_\delta$.
Set $\phi = \mathsf{CP}_\delta(\varPi \Cap X)$, then $X_\phi \subseteq X$ by the definition of $X_\phi$.
If $\phi \in \varPi'$, we are done.
So assume $\phi \notin \varPi'$.
To avoid confusion, we use $\varPi_\sim$ (resp., $\varPi_\sim'$) to denote the set $\varPi$ (resp., $\varPi'$) in the procedure, which changes over time.
Consider the iteration of the procedure which removes $\phi$ from $\varPi_\sim$ in Step~3 (such an iteration exists because $\varPi_\sim = \emptyset$ eventually).
In Step~1 of this iteration, we choose a minimal element $X^* \in \mathcal{X}$ satisfying $\varPi_\sim \Cap X^* \neq \emptyset$.
Subsequently, in Step~2, we choose $\phi$ as our $\phi^*$.
This means $\phi = \mathsf{CP}_\delta(\varPi_\sim \Cap X^*)$ at that moment, and in particular $\phi \subseteq X^*$.
It follows that $X_\phi \subseteq X^*$ by the definition of $X_\phi$.
On the other hand, we have $X^* \subseteq X_\phi$ because $X^*$ is a minimal element satisfying $\varPi_\sim \Cap X^* \neq \emptyset$ (note that $\varPi_\sim \Cap X_\phi \neq \emptyset$ as $\phi \in \varPi_\sim \Cap X_\phi$ at that moment).
Therefore, $X^* = X_\phi$.
In Step~2, we decided to \textit{not} add $\phi$ to $\varPi'_\sim$, because $\phi \notin \varPi'$ eventually.
This implies $(1+\varepsilon) \cdot |\phi|_\delta \geq |\mathsf{CP}_\delta(\varPi_\sim' \Cap X^*)|_\delta$ at that moment.
Since $\varPi_\sim' \subseteq \varPi'$ and $X^* = X_\phi$, we have $(1+\varepsilon) \cdot |\phi|_\delta \geq |\mathsf{CP}_\delta(\varPi' \Cap X_\phi)|_\delta$.
Now let us consider the range $X$.
Using the above argument and the fact $X_\phi \subseteq X$, we deduce that
\begin{equation*}
    (1+\varepsilon) \cdot |\phi|_\delta \geq |\mathsf{CP}_\delta(\varPi' \Cap X_\phi)|_\delta \geq |\mathsf{CP}_\delta(\varPi' \Cap X)|_\delta,
\end{equation*}
which immediately completes the proof.

\subsection{Proof of Lemma~\ref{lem-adjacent}}
Let $\phi,\psi \in \varPi'_{G,v,h}$ be two adjacent pairs, where $\phi = \{a,b\}$ and $\psi = \{a,c\}$.
Without loss of generality, assume $a$ is to the left of $v$ and above $h$.
Then both $b$ and $c$ are to the right of $v$ and below $h$.
Therefore, $a$ is the left-top vertex of $\mathsf{BB}(\phi \cup \psi) = \mathsf{BB}(\{a,b,c\})$, which implies that $\phi$ and $\psi$ are strongly adjacent.

\subsection{Proof of Lemma~\ref{lem-gap}}
Consider the procedure for generating $\varPi'$.
For each $\xi \in \varPi'$, we denote by $\varPi_{(\xi)}$ (resp., $\varPi'_{(\xi)}$) the set $\varPi$ (resp., $\varPi'$) just before the moment that $\xi$ is added to $\varPi'$ in Step~2 (so $\xi \in \varPi_{(\xi)}$ and $\xi \notin \varPi'_{(\xi)}$).
To prove the lemma, assume $|\phi|_\delta < |\psi|_\delta$, without loss of generality.
We first show $X_\phi \nsubseteq X_\psi$, using contradiction.
Assume $X_\phi \subseteq X_\psi$.
Consider the iteration of the procedure in which $\psi$ is added to $\varPi'$.
In Step~1 of this iteration, we select a minimal element $X^* \in \mathcal{X}$ satisfying $\varPi_{(\psi)} \Cap X^* \neq \emptyset$.
Subsequently, in Step~2, we choose $\psi$ as our $\phi^*$ (and add it to $\varPi'$).
This means $\psi = \mathsf{CP}_\delta(\varPi_{(\psi)} \Cap X^*)$, and in particular $\psi \subseteq X^*$.
We then have $X_\psi \subseteq X^*$ by the definition of $X_\psi$.
Thus, $X_\phi \subseteq X_\psi \subseteq X^*$, which implies $\phi \subseteq X^*$.
Note that either $\phi \in \varPi_{(\psi)}$ or $\phi \in \varPi'_{(\psi)}$ (because $\phi$ is eventually added to $\varPi'$).
If $\phi \in \varPi_{(\psi)}$, then $\phi \in \varPi_{(\psi)} \Cap X^*$, and hence $\psi \neq \mathsf{CP}_\delta(\varPi_{(\psi)} \Cap X^*)$ by our assumption $|\phi|_\delta < |\psi|_\delta$, which results in a contradiction.
In the case that $\phi \in \varPi'_{(\psi)}$, we have $\phi \in \varPi'_{(\psi)} \Cap X^*$, which implies $(1+\varepsilon) \cdot |\psi|_\delta > |\psi|_\delta > |\mathsf{CP}_\delta(\varPi'_{(\psi)} \Cap X^*)|_\delta$.
But $\psi$ can be added to $\varPi'$ only if $\varPi'_{(\psi)} \Cap X^* = \emptyset$ (which is not true as $\phi \in \varPi'_{(\psi)} \Cap X^*$) or $(1+\varepsilon) \cdot |\psi|_\delta < |\mathsf{CP}_\delta(\varPi'_{(\psi)} \Cap X^*)|_\delta$.
So this case also results in a contradiction.
Therefore, $X_\phi \nsubseteq X_\psi$.
This implies $X_\psi \subseteq X_\phi$, because $\phi,\psi$ are strongly adjacent and $\mathcal{X}$ is well-behaved on $\varPi$.
In fact, $X_\psi \subsetneq X_\phi$, since $X_\phi \nsubseteq X_\psi$.

Using this observation, we further show that $|\psi|_\delta > (1+\varepsilon) \cdot |\phi|_\delta$.
Consider the iteration in which $\phi$ is added to $\varPi'$.
In Step~1 of this iteration, we select a minimal element $X^* \in \mathcal{X}$ satisfying $\varPi_{(\phi)} \Cap X^* \neq \emptyset$.
Then in Step~2, we choose $\phi$ as our $\phi^*$ (and add it to $\varPi'$).
Therefore, $\phi = \mathsf{CP}_\delta(\varPi_{(\phi)} \Cap X^*)$, and in particular $\phi \subseteq X^*$.
We observe that $\psi \notin \varPi_{(\phi)}$.
Indeed, we have $X_\psi \subsetneq X_\phi \subseteq X^*$ where the second ``$\subseteq$'' follows from the fact $\phi \subseteq X^*$.
But $X^* \in \mathcal{X}$ is a minimal element satisfying $\varPi_{(\phi)} \Cap X^* \neq \emptyset$, hence $\varPi_{(\phi)} \Cap X_\psi = \emptyset$ and $\psi \notin \varPi_{(\phi)}$.
This further implies $\psi \in \varPi'_{(\phi)}$ (because $\psi$ is eventually in $\varPi'$).
Note that $\phi$ can be added to $\varPi'$ only if $\varPi'_{(\phi)} = \emptyset$ or $(1+\varepsilon) \cdot |\phi|_\delta < |\mathsf{CP}_\delta(\varPi'_{(\phi)} \Cap X^*)|_\delta$, according to the conditions in Step~2.
The former is not true as $\psi \in \varPi'_{(\phi)}$, so the latter must be true.
Furthermore, we observe that $\psi \subseteq X_\psi \subsetneq X_\phi \subseteq X^*$, and hence $\psi \in \varPi_{(\phi)}' \Cap X^*$.
Thus, $(1+\varepsilon) \cdot |\phi|_\delta < |\mathsf{CP}_\delta(\varPi'_{(\phi)} \Cap X^*)|_\delta \leq |\psi|_\delta$.

\subsection{Proof of Theorem~\ref{thm-coreset}}
The proof is a direct generalization of that for $d=2$.
Again we define $f(m) = \sup_{G \in \mathcal{G}_m} |\varPi'_G|$, and our goal is to show $f(m) = O(\varepsilon^{-1} m\log^dm)$.
Recall that in the case of $d=2$, we get $f(m) \leq 2f(m/2)+f'(m)$ by separating the points vertically and $f'(m) \leq 2f'(m/2)+f''(m)$ by further separating the points horizontally.
Here we do the same thing, but we need to do it $d$ times.
We say a pair $\{a,b\}$ of points in $\mathbb{R}^d$ \textit{crosses} a hyperplane $h$ if $a$ and $b$ are on different sides of $h$.
If $h_1,\dots,h_i$ are hyperplanes in $\mathbb{R}^d$, we use the notation $\varPi'_{G,h_1,\dots,h_i}$ to denote the subset of $\varPi'_G$ consisting of all the pairs that cross $h_1,\dots,h_i$ simultaneously.
Let $H_i$ be the set of all hyperplanes in $\mathbb{R}^d$ parallel to the hyperplane $x_i=0$, for $i \in [d]$.
Define $f_i(m) = \sup_{G \in \mathcal{G},h_1 \in H_1,\dots,h_i \in H_i} |\varPi'_{G,h_1,\dots,h_i}|$ for $i \in [d]$.
Then we have the recurrence $f_i(m) \leq 2f_i(m/2) + f_{i+1}(m)$ for $i \in [d-1]$ (similar to the two recurrences we achieved for $d=2$).
It suffices to bound $f_d(m)$.
By the same charging argument as we used for bounding $f''(m)$, we can deduce $f_d(m) \leq \lambda m$ where $\lambda = \lceil \log_{1+\varepsilon} 2 \rceil$.
Solving the recurrences one-by-one, we finally have $f(m) = O(\varepsilon^{-1} m\log^dm)$ for all $m \in [n]$.
Therefore, $|\varPi'| = O(\lambda n\log^dn)$.

\subsection{Proof of Lemma~\ref{lem-angle}}
Since $a^*,a \in R^\textnormal{ne}$ and $L_1(a^*,o) \leq L_1(a,o)$, there exists a point $a'$ on the segment connecting $o$ and $a$ such that $L_1(a^*,o) = L_1(a',o)$.
Symmetrically, there exists a point $b'$ on the segment connecting $o$ and $b$ such that $L_1(b^*,o) = L_1(b',o)$.
By the monotonicity of $\lVert \cdot \rVert$, we have $\delta(a',b') \leq \delta(a,b)$.
Therefore, it suffices to prove $\delta(a^*,b^*) \leq (1+\varepsilon) \cdot \delta(a',b')$.
We first show $\delta(a^*,a') \leq (\varepsilon/2) \cdot \delta(a',b')$.
In the triangle $\triangle a^*oa'$, we have the relation $L_2(a^*,a')/L_2(a',o) = \sin \angle a^*oa'/\sin \angle oa^*a'$, where $L_2(\cdot,\cdot)$ denotes the Euclidean metric.
Note that $\angle a^*oa' = \angle a^*oa \leq \theta$, and hence $\sin \angle a^*oa' \leq \theta$.
On the other hand, we have $\pi/4 \leq \angle oa^*a' \leq 3\pi/4$ because $a^*$ is to the northeast of $o$ and $\mathsf{x}(a^*)+\mathsf{y}(a^*) = \mathsf{x}(a')+\mathsf{y}(a')$, where the latter equation follows from the fact $L_1(a^*,o) = L_1(a',o)$.
Therefore, $\sin \angle oa^*a' \geq 1/\sqrt{2}$ and $L_2(a^*,a')/L_2(a',o) \leq \sqrt{2} \theta$.
Applying Lemma~\ref{lem-mono}, we have
\begin{equation*}
    \frac{\delta(a^*,a')}{\delta(a',o)} \leq \frac{2 L_2(a^*,a') \cdot \max\{\lVert e_1 \rVert,\lVert e_2 \rVert\}}{(1/\sqrt{2}) \cdot L_2(a',o) \cdot \min\{\lVert e_1 \rVert,\lVert e_2 \rVert\}} = \frac{2 \sqrt{2} \cdot L_2(a^*,a')}{L_2(a',o)} \leq 4 \theta = \varepsilon/2.
\end{equation*}
Now we see $\delta(a^*,a') \leq (\varepsilon/2) \cdot \delta(a',o)$.
By the monotonicity of $\lVert \cdot \rVert$, we have $\delta(a',b') \geq \delta(a',o)$ and thus $\delta(a^*,a') \leq (\varepsilon/2) \cdot \delta(a',b')$.
Using the same argument, we can also deduce $\delta(b^*,b') \leq (\varepsilon/2) \cdot \delta(a',b')$.
By the triangle inequality, we then have $\delta(a^*,b^*) \leq \delta(a^*,a') + \delta(a',b') + \delta(b^*,b') \leq (1+\varepsilon) \cdot \delta(a',b')$, which completes the proof since $\delta(a',b') \leq \delta(a,b)$.

\subsection{Proof of Lemma~\ref{lem-correct}}
Clearly, $\mathsf{Bich}_o(T) \subseteq \varPi_1 \Cap R$ since all pairs in $\mathsf{Bich}_o(T)$ are $o$-anchored NE-SW pairs and are contained in $R$.
Thus, $\phi_1 \in \varPi_1 \Cap R$.
To see $|\phi_1|_\delta \leq (1+\varepsilon) \cdot |\mathsf{CP}_\delta(\varPi_1 \Cap R)|_\delta$, let $\{a,b\} = \mathsf{CP}_\delta(\varPi_1 \Cap R)$.
Suppose $a \in U_i$ and $b \in V_j$.
We claim that there exist $a^* \in \{a_i,a_i'\}$ and $b^* \in \{b_j,b_j'\}$ such that \textbf{(i)} $\text{cl}(a^*) \neq \text{cl}(b^*)$ and \textbf{(ii)} $L_1(a^*,o) \leq L_1(a,o)$ and $L_1(b^*,o) \leq L_1(b,o)$.
We consider four cases.
The first case is $\text{cl}(a) = \text{cl}(a_i)$ and $\text{cl}(b) = \text{cl}(b_j)$.
In this case, we let $a^* = a_i$ and $b^* = b_j$.
Then $\text{cl}(a^*) = \text{cl}(a) \neq \text{cl}(b) = \text{cl}(b^*)$.
Also, we have $L_1(a^*,o) \leq L_1(a,o)$ and $L_1(b^*,o) \leq L_1(b,o)$ by the definition of $a_i$ and $b_j$.
Thus the claim holds.
The second case is $\text{cl}(a) = \text{cl}(a_i)$ and $\text{cl}(b) \neq \text{cl}(b_j)$.
In this case, we let $a^* = a_i$ and choose $b^* \in \{b_j,b_j'\}$ such that $\text{cl}(a^*) \neq \text{cl}(b^*)$; note that this is possible because $\text{cl}(b_j) \neq \text{cl}(b_j')$.
We have $L_1(a^*,o) \leq L_1(a,o)$ by the definition of $a_i$.
Since $\text{cl}(b) \neq \text{cl}(b_j)$, we have $L_1(b_j,o) \leq L_1(b_j',o) \leq L_1(b,o)$ by the definition of $b_j'$, which implies $L_1(b^*,o) \leq L_1(b,o)$.
Thus the claim holds.
The third case $\text{cl}(a) \neq \text{cl}(a_i)$ and $\text{cl}(b) = \text{cl}(b_j)$ is symmetric to the second one.
The last case is $\text{cl}(a) \neq \text{cl}(a_i)$ and $\text{cl}(b) \neq \text{cl}(b_j)$.
In this case, we have $L_1(a_i,o) \leq L_1(a_i',o) \leq L_1(a,o)$ and $L_1(b_j,o) \leq L_1(b_j',o) \leq L_1(b,o)$ by the definition of $a_i'$ and $b_j'$.
So the property \textbf{(ii)} holds for all $a^* \in \{a_i,a_i'\}$ and all $b^* \in \{b_j,b_j'\}$.
We only need to choose $a^* \in \{a_i,a_i'\}$ and $b^* \in \{b_j,b_j'\}$ such that $\text{cl}(a^*) \neq \text{cl}(b^*)$; this is possible because $\text{cl}(a_i) \neq \text{cl}(a_i')$ and $\text{cl}(b_j) \neq \text{cl}(b_j')$.
Thus, the claim holds.
Using the claim, we can prove the lemma as follows.
Set $\phi^* = \{a^*,b^*\}$.
We have $\phi^* \in \mathsf{Bich}(T)$ because $a^*,b^* \in T$ and $\text{cl}(a^*) \neq \text{cl}(b^*)$.
Furthermore, $\phi^*$ is $o$-anchored because $a^* \in U_i$ and $b^* \in V_j$.
Therefore, $\phi^* \in \mathsf{Bich}_o(T)$.
As $a^*, a \in U_i$ and $b^*, b \in V_j$, we have $\angle a^* o a \leq \theta$ and $\angle b^* o b \leq \theta$.
In addition, $L_1(a^*,o) \leq L_1(a,o)$ and $L_1(b^*,o) \leq L_1(b,o)$ by the property \textbf{(ii)} above.
Thus we have $|\phi^*|_\delta = \delta(a^*,b^*) \leq (1+\varepsilon) \cdot \delta(a,b)$ by Lemma~\ref{lem-angle}.
As a result, $|\phi_1|_\delta = |\mathsf{CP}_\delta(\mathsf{Bich}_o(T))|_\delta \leq |\phi^*|_\delta \leq (1+\varepsilon) \cdot \delta(a,b)$.

\subsection{Proof of Lemma~\ref{lem-top2ds}}
In the range-search problem defined in the lemma, the query ranges are of the form $\bigcap_{i=1}^c H_i^\dagger$ where $H_i^\dagger$ is a translate of $H_i$, and the target of searching is the two lightest points with different colors (or the \textit{top-2} points for short) in the query range.
We first notice this is a decomposable range-search problem.
Indeed, if $A = A_1 \cup A_2$, then the top-2 points in $A \cap (\bigcap_{i=1}^c H_i^\dagger)$ can be directly computed from the top-2 points in $A_1 \cap (\bigcap_{i=1}^c H_i^\dagger)$ and the top-2 points in $A_2 \cap (\bigcap_{i=1}^c H_i^\dagger)$.
To design the data structure, we use induction on $c$.
The base case is $c=2$ will be handled later.
Suppose the desired data structure exists for $c-1$ halfplanes, we consider the case in which there are $c$ halfplanes $H_1,\dots,H_c$ and we want to build the desired data structure for $H_1,\dots,H_c$.
Without loss of generality, we may assume $H_c$ is the halfplane $x \geq 0$ (if not, we can rotate the dataset $A$ as well as the halfplanes).
We store $A$ in a standard 1D range tree $\mathbf{T}$ built on the $x$-coordinates.
Basically, each leaf of $\mathbf{T}$ stores one point in $A$; if a leaf $\mathbf{l}$ is to the left of another leaf $\mathbf{l}'$ in $\mathbf{T}$, then the point stored at $\mathbf{l}$ has a smaller $x$-coordinate than the point stored at $\mathbf{l}'$.
Each node $\mathbf{u} \in \mathbf{T}$ correspond to a canonical subset $A(\mathbf{u}) \subseteq A$ which is by definition the set of the points stored in the subtree of $\mathbf{u}$ (and thus the points in $A(\mathbf{u})$ have consecutive $x$-coordinates in $A$).
At each node $\mathbf{u} \in \mathbf{T}$, we store the top-2 data structure built on $A(\mathbf{u})$ for the $c-1$ halfplanes $H_1,\dots,H_{c-1}$; by our induction hypothesis, we have such a data structure using $O(|A(\mathbf{u})| \log^{c-2} |A(\mathbf{u})|)$ space and $O(\log^{c-2} |A(\mathbf{u})|)$ query time.
The range tree $\mathbf{T}$ together with the data structures stored at each node is just the desired top-2 data structure, which clearly uses $O(m \log^{c-1} m)$ space.
Consider a query $H_1^\dagger,\dots,H_c^\dagger$ where $H_i^\dagger$ is a translate of $H_i$.
Suppose the equation of $H_c^\dagger$ is $x \geq \mu$.
We first find the canonical nodes in $\mathbf{T}$ corresponding to the range $[\mu,\infty)$, say $\mathbf{u}_1,\dots,\mathbf{u}_t$ where $t = O(\log m)$.
This can be done $O(\log n)$ time.
By the property of the canonical nodes, $A \cap H_c^\dagger = \bigcup_{j=1}^t A(\mathbf{u}_j)$.
For each $j \in [t]$, we query the data structure associated to $\mathbf{u}_j$ with the halfplanes $H_1^\dagger,\dots,H_{c-1}^\dagger$; the answer returned is the top-2 points in $A(\mathbf{u}_j) \cap (\bigcap_{i=1}^{c-1} H_i^\dagger)$.
We collect all the answers (which are $2t$ points), and then compute the top-2 points among these $2t$ points, which are just the top-2 points in $A \cap (\bigcap_{j=1}^c H_j^\dagger)$.
The overall query time is $O(\log^{c-1} n)$ because the query time of each associated data structure is bounded by $O(\log^{c-2} n)$.

The remaining task is to handle the base case $c=2$.
Let $H_1: \alpha_1 x + \beta_1 y \geq 0$ and $H_2: \alpha_2 x + \beta_2 y \geq 0$ be the halfplanes.
Consider the map $\pi: \mathbb{R}^2 \rightarrow \mathbb{R}^2$ defined as $(x,y) \mapsto (\alpha_1 x + \beta_1 y, \alpha_2 x + \beta_2 y)$.
Given two halfplanes $H_1^\dagger: \alpha_1 x + \beta_1 y \geq \mu_1$ and $H_2^\dagger: \alpha_2 x + \beta_2 y \geq \mu_2$, we have $a \in H_1^\dagger \cap H_2^\dagger$ iff $\pi(a) \in [\mu_1,\infty) \times [\mu_2,\infty)$.
Using this observation, we reduce the task to building a top-2 data structure for (northeast) quadrant queries.
Again, we store $A$ in a 1D range tree $\mathbf{T}$ built on the $x$-coordinates.
At each node $\mathbf{u} \in \mathbf{T}$, we store a (balanced) binary search tree $\mathbf{T}_\mathbf{u}$ built on the canonical subset $A(\mathbf{u})$ which uses the $y$-coordinates of the points as keys.
Then each node $\mathbf{v} \in \mathbf{T}_\mathbf{u}$ corresponds to a point $a_\mathbf{v} \in A(\mathbf{u})$; we define $A_{\geq \mathbf{v}}(\mathbf{u}) = \{a \in A(\mathbf{u}): \mathsf{y}(a) \geq \mathsf{y}(a_\mathbf{v})\}$.
We associate to each node $\mathbf{v} \in \mathbf{T}_\mathbf{u}$ the top-2 points in $A_{\geq \mathbf{v}}(\mathbf{u})$.
The space cost of each binary search tree $\mathbf{T}_\mathbf{u}$ is $O(|A(\mathbf{u})|)$, and thus the space cost of $\mathbf{T}$ is $O(m \log m)$.
Consider a query quadrant $Q = [x^-,\infty) \times [y^-,\infty)$.
We first find the canonical nodes in $\mathbf{T}$ corresponding to the range $[x^-,\infty)$, say $\mathbf{u}_1,\dots,\mathbf{u}_t$ where $t = O(\log m)$.
For each $i \in [t]$, we find the smallest node $\mathbf{v}_i \in \mathbf{T}_{\mathbf{u}_i}$ whose key is at least $y^-$, then $A(\mathbf{u}_i) \cap Q = A_{\geq \mathbf{v}_i}(\mathbf{u}_i)$.
We then collect the top-2 points in $A_{\geq \mathbf{v}_1}(\mathbf{u}_1),\dots,A_{\geq \mathbf{v}_t}(\mathbf{u}_t)$, which are associated to the nodes $\mathbf{v}_1,\dots,\mathbf{v}_t$, and compute the top-2 points among these $2t$ points, which are just the top-2 points in $A \cap Q$.
Now the overall query time is $O(\log^2 m)$, as searching for each $\mathbf{v}_i$ in $\mathbf{T}_{\mathbf{u}_i}$ takes $O(\log m)$ time.
To achieve $O(\log m)$ query time, we apply the \textit{fractional cascading} technique.
Specifically, for each pair $(\mathbf{u},\mathbf{u}')$ of nodes in $\mathbf{T}$ where $\mathbf{u}'$ is a child of $\mathbf{u}$, we store at each node $\mathbf{v} \in \mathbf{T}_\mathbf{u}$ a pointer connecting to the smallest node $\mathbf{v}' \in \mathbf{T}_{\mathbf{u}'}$ whose key is greater than or equal to the key of $\mathbf{v}$.
Storing these pointers does not increase the asymptotic space cost of the data structure.
Taking advantage of these pointers, when we find the canonical nodes $\mathbf{u}_1,\dots,\mathbf{u}_t$, we can also find $\mathbf{v}_1,\dots,\mathbf{v}_t$ at the same time (we omit the details here as it is just the standard fractional cascading technique), and then the overall query time becomes $O(\log m)$.

\subsection{Proof of Theorem~\ref{thm-anchored3d}}
The idea to design the desired data structure is the same as that in $\mathbb{R}^2$.
Let $\lVert \cdot \rVert$ be the monotone norm on $\mathbb{R}^3$ inducing $\delta$.
Again, we assume $\lVert (1,0,0) \rVert = \lVert (0,1,0) \rVert = \lVert (0,0,1) \rVert = 1$, without loss of generality.
Consider a query $(B,o)$ where $B = [x^-,x^+] \times [y^-,y^+] \times [z^-,z^+] \in \mathcal{B}$ and $o \in \mathbb{R}^3$.
If $o \notin B$, then $\mathsf{Bich}_o(S) \Cap B = \emptyset$ and nothing should be reported.
So suppose $o \in B$.
Recall that in $\mathbb{R}^2$, we reduce the task to considering only the pairs with one point in $R^\text{ne}$ and one point in $R^\text{sw}$.
What we do here is similar.
Let $B^+ = [\mathsf{x}(o),x^+] \times [\mathsf{y}(o),y^+] \times [\mathsf{z}(o),z^+]$ and $B^- = [x^-,\mathsf{x}(o)] \times [y^-,\mathsf{y}(o)] \times [z^-,\mathsf{z}(o)]$.
Define $\varPi \subseteq \mathsf{Bich}_o(S)$ as the subset consisting of all pairs with one point in $B^-$ and one point in $B^+$.
It suffices to show how to compute a $(1+\varepsilon)$-ACP in $\varPi \Cap B$.
Set $\theta = \varepsilon/18$.
Similarly to Lemma~\ref{lem-angle}, we have the following observation.
\begin{lemma} \label{lem-angle3d}
Let $a^*,a \in B^+$ such that $\angle a^*oa \leq \theta$ and $L_1(a^*,o) \leq L_1(a,o)$.
Let $b^*,b \in B^-$ such that $\angle b^*ob \leq \theta$ and $L_1(b^*,o) \leq L_1(b,o)$.
Then $\delta(a^*,b^*) \leq (1+\varepsilon) \cdot \delta(a,b)$.
\end{lemma}
\textit{Proof.}
The proof is similar to that of Lemma~\ref{lem-angle}.
Since $a^*,a \in R^\textnormal{ne}$ and $L_1(a^*,o) \leq L_1(a,o)$, there exists a point $a'$ on the segment connecting $o$ and $a$ such that $L_1(a^*,o) = L_1(a',o)$.
Symmetrically, there exists a point $b'$ on the segment connecting $o$ and $b$ such that $L_1(b^*,o) \leq L_1(b',o)$.
By the monotonicity of $\lVert \cdot \rVert$, we have $\delta(a',b') \leq \delta(a,b)$.
Therefore, it suffices to prove that $\delta(a^*,b^*) \leq (1+\varepsilon) \cdot \delta(a',b')$.
We first show $\delta(a^*,a') \leq (\varepsilon/2) \cdot \delta(a',b')$.
In the triangle $\triangle a^*oa'$, we have the relation $L_2(a^*,a')/L_2(a',o) = \sin \angle a^*oa'/\sin \angle oa^*a'$, where $L_2(\cdot,\cdot)$ denotes the Euclidean metric.
Note that $\angle a^*oa' = \angle a^*oa \leq \theta$, and hence $\sin \angle a^*oa' \leq \theta$.
On the other hand, one can easily verify that $\sin \angle oa^*a' \geq 1/\sqrt{3}$ because $a^* \in B^+$ and $\mathsf{x}(a^*)+\mathsf{y}(a^*)+\mathsf{z}(a^*) = \mathsf{x}(a')+\mathsf{y}(a')+\mathsf{z}(a')$, where the latter equation follows from the fact $L_1(a^*,o) = L_1(a',o)$.
Thus, $L_2(a^*,a')/L_2(a',o) \leq \sqrt{3} \theta$.
Applying Lemma~\ref{lem-mono}, we have
\begin{equation*}
    \frac{\delta(a^*,a')}{\delta(a',o)} \leq \frac{3 L_2(a^*,a') \cdot \max\{\lVert e_1 \rVert,\lVert e_2 \rVert,\lVert e_3 \rVert\}}{(1/\sqrt{3}) \cdot L_2(a',o) \cdot \min\{\lVert e_1 \rVert,\lVert e_2 \rVert,\lVert e_3 \rVert\}} = \frac{3 \sqrt{3} \cdot L_2(a^*,a')}{L_2(a',o)} \leq 9 \theta = \varepsilon/2.
\end{equation*}
Now we see $\delta(a^*,a') \leq (\varepsilon/2) \cdot \delta(a',o)$.
By the monotonicity of $\lVert \cdot \rVert$, we have $\delta(a',b') \geq \delta(a',o)$ and thus $\delta(a^*,a') \leq (\varepsilon/2) \cdot \delta(a',b')$.
Using the same argument, we can also deduce $\delta(b^*,b') \leq (\varepsilon/2) \cdot \delta(a',b')$.
By the triangle inequality, we then have $\delta(a^*,b^*) \leq \delta(a^*,a') + \delta(a',b') + \delta(b^*,b') \leq (1+\varepsilon) \cdot \delta(a',b')$, which completes the proof since $\delta(a',b') \leq \delta(a,b)$.
\hfill $\Box$
\smallskip

\noindent
Based on the above lemma, the idea for solving the problem is similar to that in $\mathbb{R}^2$, namely, dividing $B^+$ and $B^-$ into small regions.
Set $k = \lceil \pi/4\theta \rceil$.
Define $2(k+1)$ planes $\varGamma_0,\dots,\varGamma_k$ and $\varGamma_0',\dots,\varGamma_k'$ as
\begin{equation*}
\begin{aligned}
    \varGamma_i:\ & (x-\mathsf{x}(o)) \cdot \cos(i\pi/2k) - (y-\mathsf{y}(o)) \cdot \sin(i\pi/2k) = 0, \\
    \varGamma_i':\ & ((x-\mathsf{x}(o))+(y-\mathsf{y}(o))) \cdot \cos(i\pi/2k) - \sqrt{2} (z-\mathsf{z}(o)) \cdot \sin(i\pi/2k) = 0.
\end{aligned}
\end{equation*}
Clearly, these planes divide $B^+$ into $k^2$ regions.
For each $i \in [k]$ and each $j \in [k]$, there is exactly one region bounded by $\varGamma_{i-1},\varGamma_i$ and $\varGamma_{j-1}',\varGamma_j$, which we denote by $U_{i,j}$.
Each $U_{i,j}$ can be a 5-polyhedron, 6-polyhedron, or 7-polyhedron (i.e., a polyhedron of 5, 6, or 7 faces).
Due to the choice of $\varGamma_0,\dots,\varGamma_k$ and $\varGamma_0',\dots,\varGamma_k'$, for any $a,b \in U_{i,j}$, we have $\angle aob \leq \theta$.
Similarly, the planes also divide $B^-$ into $k^2$ regions, and we denote by $V_{i,j}$ the region bounded by $\varGamma_{i-1},\varGamma_i$ and $\varGamma_{j-1}',\varGamma_j$.
Again, for any $a,b \in V_{i,j}$, we have $\angle aob \leq \theta$.
In order to compute a $(1+\varepsilon)$-ACP in $\varPi \Cap B$, we use the same approach as that in $\mathbb{R}^2$.
For each $i \in [k]$ and each $j \in [k]$, we report two points $a_{i,j},a_{i,j}' \in S \cap U_{i,j}$, where $a_{i,j}$ is the $L_1$-nearest-neighbor of $o$ among the points in $S \cap U_{i,j}$ and $a_{i,j}'$ is the $L_1$-nearest-neighbor of $o$ among the points in $S \cap U_{i,j}$ whose colors are different from $\text{cl}(a_{i,j})$.
Also, for each $i \in [k]$ and each $j \in [k]$, we report two points $b_{i,j},b_{i,j}' \in S \cap V_{i,j}$, where $b_{i,j}$ is the $L_1$-nearest-neighbor of $o$ among the points in $S \cap V_{i,j}$ and $b_{i,j}'$ is the $L_1$-nearest-neighbor of $o$ among the points in $S \cap V_{i,j}$ whose colors are different from $\text{cl}(b_{i,j})$.
Let $T = \{a_{i,j},a_{i,j}',b_{i,j},b_{i,j}': i,j \in [k]\}$.
We then compute $\phi = \mathsf{CP}_\delta(\mathsf{Bich}_o(T))$ by brute-force in $O(k^4)$ time (note that $|T| = 4k^2$).
By applying Lemma~\ref{lem-angle3d}, we can prove that $\phi$ is indeed a $(1+\varepsilon)$-ACP in $\varPi \Cap B$.
\begin{lemma}
We have $\phi \in \varPi \Cap B$ and $|\phi|_\delta \leq (1+\varepsilon) \cdot |\mathsf{CP}_\delta(\varPi \Cap B)|_\delta$.
\end{lemma}
\textit{Proof.}
Same as the proof of Lemma~\ref{lem-correct}.
\hfill $\Box$
\smallskip

\noindent
Now we show how to efficiently compute $a_{i,j},a_{i,j}'$ and $b_{i,j},b_{i,j}'$ for $i,j \in [k]$.
It suffices to consider the points $a_{i,j},a_{i,j}'$.
As mentioned before, each $U_{i,j}$ can be a 5-polyhedron, 6-polyhedron, or 7-polyhedron.
For simplicity, here we do not distinguish these cases (unlike what we do in $\mathbb{R}^2$).
Instead, for each $i \in [k]$ and $j \in [k]$, we shall build a single data structure $\mathbf{C}_{i,j}$ for computing $a_{i,j},a_{i,j}'$.
To this end, we generalize Lemma~\ref{lem-top2ds} to $\mathbb{R}^3$, which is straightforward.
\begin{lemma} \label{lem-top2ds3d}
Let $c \geq 2$ be a constant and $H_1,\dots,H_c$ halfspaces in $\mathbb{R}^3$.
Given a weighted colored dataset $A$ in $\mathbb{R}^3$ of size $m$, one can build an $O(m \log^{c-1} m)$-space data structure on $A$ that can report in $O(\log^{c-1} m)$ time, for any specified $c$ halfspaces $H_1^\dagger,\dots,H_c^\dagger$ where $H_i^\dagger$ is a translate of $H_i$, two points $a$ and $a'$ where $a$ is the lightest point in $A \cap (\bigcap_{i=1}^c H_i^\dagger)$ and $a'$ is the lightest point in $A \cap (\bigcap_{i=1}^c H_i^\dagger)$ whose color is different from $\textnormal{cl}(a)$.
\end{lemma}
\textit{Proof.}
Same as the proof of Lemma~\ref{lem-top2ds}.
\hfill $\Box$
\smallskip

\noindent
Note that for a point $a \in B^+$, the smaller $\mathsf{x}(a)+\mathsf{y}(a)+\mathsf{z}(a)$ is, the smaller $L_1(a,o)$ is.
As such, we assign each point $a \in S$ a \textit{weight} equal to $\mathsf{x}(a)+\mathsf{y}(a)+\mathsf{z}(a)$.
Then $a_{i,j}$ is just the lightest point in $S \cap U_{i,j}$ and $a_{i,j}'$ is the lightest point in $S \cap U_{i,j}$ whose color is different from $\text{cl}(a_{i,j})$.
This allows us to apply the above lemma.
Fix $i,j \in [k]$.
We define 7 halfspaces $H_1: x \leq 0$, $H_2: y \leq 0$, $H_3: z \leq 0$,
\begin{equation*}
\begin{aligned}
    H_4:\ & x \cdot \cos((i-1)\pi/2k) - y \cdot \sin((i-1)\pi/2k) \geq 0, \\
    H_5:\ & x \cdot \cos(i\pi/2k) - y \cdot \sin(i\pi/2k) \leq 0, \\
    H_6:\ & (x+y) \cdot \cos((i-1)\pi/2k) - z \cdot \sqrt{2}\sin((i-1)\pi/2k) \geq 0, \\
    H_7:\ & (x+y) \cdot \cos(i\pi/2k) - z \cdot \sqrt{2}\sin(i\pi/2k) \leq 0.
\end{aligned}
\end{equation*}
Then $U_{i,j}$ is always the intersection of 7 halfspaces $H_1^\dagger,\dots,H_7^\dagger$ where $H_i^\dagger$ is a translate of $H_i$.
Thus, by Lemma~\ref{lem-top2ds3d}, we can build the data structure $\mathbf{C}_{i,j}$ with $O(n \log^6 n)$ space and $O(\log^6 n)$ query time.
The space cost of all these data structures is then $O(k^2 n \log^6 n)$, and the total time for computing all $a_{i,j},a_{i,j}'$ is $O(k^2 \log^6 n)$.
The points $b_{i,j},b_{i,j}'$ can be computed by building similar data structures.
Including the $O(k^4)$ time for finding $\mathsf{CP}_\delta(\mathsf{Bich}_o(T))$, the time for computing $\phi$ is $O(k^2 \log^6 n + k^4)$, which is also the query time of our anchored CRCP data structure.
Since $k = O(\varepsilon^{-1})$, the theorem is proved.

\subsection{Proof of Lemma~\ref{lem-xue}}
This lemma follows immediately from a discussion in \cite{xue2018new}, though it was not explicitly claimed in \cite{xue2018new}.
The method is to reduce the range-search problem to a point-location problem in a linear-complexity planar subdivision.
Please see \cite{xue2018new} (Section~2 and 3) for details.
We remark that in \cite{xue2018new} the weight of a pair $\phi$ is always equal to $|\phi|_{L_2}$.
However, the method clearly works for arbitrarily-weighted set of point pairs.

\subsection{Proof of Lemma~\ref{lem-candOn}}
Let $\phi = \{a,b\} \in \varPi_2'$ where $a$ is the southwest point of $\phi$.
We claim that $b$ is the $\delta$-nearest-neighbor of $a$ among all the points to the northeast of $a$ whose colors are different from $\text{cl}(a)$.
To see this, suppose $Q \in \mathcal{Q}^\text{ne}$ is a northeast quadrant such that $\phi = \mathsf{CP}_\delta(\varPi_2 \Cap Q)$.
Then $Q$ contains all the points in $S$ that are to the northeast of $a$.
Therefore, $\varPi_2 \Cap Q$ contains all the bichromatic pairs that consist of $a$ and another point to the northeast of $a$.
As such, our claim holds.
Using this claim, if we charge every pair in $\varPi_2'$ to its southwest point, then every point in $S$ is charged at most once, which implies $|\varPi_2'| \leq n$.

\subsection{Proof of Lemma~\ref{lem-rect1correct}}
It is clear that $\varPhi_1 \subseteq \mathsf{Bich}(S) \Cap R$, because every query we made returns a bichromatic pair of points in $S \cap R$.
Thus, $\mathsf{CP}_\delta(\varPhi_1) \in \mathsf{Bich}(S) \Cap R$.
To see $|\mathsf{CP}_\delta(\varPhi_1)|_\delta \leq (1+\varepsilon) \cdot |\mathsf{CP}_\delta(\mathsf{Bich}(S) \Cap R)|_\delta$, let $\phi^* = \mathsf{CP}_\delta(\mathsf{Bich}(S) \Cap R)$.
It suffices to show that $|\phi|_\delta \leq (1+\varepsilon) \cdot |\phi^*|_\delta$ for some $\phi \in \varPhi_1$.
If $\phi^* \subseteq S^\text{ne}(\mathbf{v})$, then the pair $\phi$ returned by $\mathbf{A}^\text{ne}(\mathbf{v})$ satisfies the desired property.
The cases $\phi^* \subseteq S^\text{se}(\mathbf{v})$, $\phi^* \subseteq S^\text{sw}(\mathbf{v})$, $\phi^* \subseteq S^\text{nw}(\mathbf{v})$ are similar.
If $\phi^* \subseteq S(\mathbf{u}'_i)$ for some $i \in [t']$, then the pair $\phi$ returned by $\mathbf{B}(\mathbf{u}'_i)$ satisfies the property.
If $\phi^* \subseteq S(\mathbf{u}''_i)$ for some $i \in [t'']$, then the pair $\phi$ returned by $\mathbf{B}(\mathbf{u}''_i)$ satisfies the property.
Finally, if all of the above conditions are not true, then $\phi^*$ must be $o$-anchored for some $o \in O_1$.
Indeed, $\phi^*$ must cross $l_\mathbf{u}$ or $l_\mathbf{v}$, say $\phi^*$ crosses $l_\mathbf{u}$.
Assume the two points of $\phi^*$ belong to $S(\mathbf{u}''_i)$ and $S(\mathbf{u}''_j)$ respectively where $i<j$.
Then $\phi^*$ is $o$-anchored for $o = o''_i \in O_1$ (recall that $o''_i$ is the point on $l_\mathbf{u}$ with $y$-coordinate $y_i$).
Now the pair $\phi$ returned by $\mathbf{C}$ for the query $(R,o)$ satisfies the property.
As a result, $|\mathsf{CP}_\delta(\varPhi_1)|_\delta \leq (1+\varepsilon) \cdot |\phi^*|_\delta$.

\subsection{Proof of Lemma~\ref{lem-rect2correct}}
It is clear that $\varPhi_2 \subseteq \mathsf{Bich}(S) \Cap R$, because every query we made above returns a bichromatic pair of points in $S \cap R$.
Thus, $\mathsf{CP}_\delta(\varPhi_2) \in \mathsf{Bich}(S) \Cap R$.
To see $|\mathsf{CP}_\delta(\varPhi_2)|_\delta \leq (1+\varepsilon) \cdot |\mathsf{CP}_\delta(\mathsf{Bich}(S) \Cap R)|_\delta$, let $\phi^* = \mathsf{CP}_\delta(\mathsf{Bich}(S) \Cap R)$.
It suffices to show that $|\phi|_\delta \leq (1+\varepsilon) \cdot |\phi^*|_\delta$ for some $\phi \in \varPhi_2$.
If $\phi^* \subseteq S(\mathbf{u}'_i)$ for some $i \in [t']$, then the pair $\phi$ returned by $\mathbf{B}(\mathbf{u}'_i)$ satisfies the property.
If $\phi^* \subseteq S(\mathbf{u}''_i)$ for some $i \in [t'']$, then the pair $\phi$ returned by $\mathbf{B}(\mathbf{u}''_i)$ satisfies the property.
If both of the above conditions are not true, then $\phi^*$ must be $o$-anchored for some $o \in O_2$.
Indeed, assume the two points of $\phi^*$ belong to $S(\mathbf{u}'_{i'})$ and $S(\mathbf{u}'_{j'})$ respectively where $i'<j'$.
Also, assume the two points of $\phi^*$ belong to $S(\mathbf{u}''_{i''})$ and $S(\mathbf{u}''_{j''})$ respectively where $i''<j''$.
Then $\phi^*$ is $o$-anchored for $o = o_{i',i''} \in O_2$ (recall that $o_{i',i''} = (x_{i'},y_{i''})$).
In this case, the pair $\phi$ returned by $\mathbf{C}$ for the query $(R,o)$ satisfies the property.
As a result, $|\mathsf{CP}_\delta(\varPhi_2)|_\delta \leq (1+\varepsilon) \cdot |\phi^*|_\delta$.

\subsection{Proof of Lemma~\ref{lem-xue'}}
The lemma is immediately implied by Lemma~\ref{lem-xue}.
Define a map $\pi: \mathbb{R}^d \rightarrow \mathbb{R}^2$ defined as $(x_1,\dots,x_d) \mapsto (x_1,x_2)$.
Let $\pi(\varPi') = \{\pi(\phi): \phi \in \varPi'\}$ and we regard $\pi(\varPi')$ as a weighted set by letting $\pi(\phi) \in \pi(\varPi')$ have the same weight as $\phi \in \varPi'$.
We first consider the slab queries $\mathcal{L}$.
Clearly, it suffices to handle the queries in the sub-collection $\mathcal{L}' \subseteq \mathcal{L}$ consisting of slabs of the form $[x^-,x^+] \times \mathbb{R}^{d-1}$ (i.e., bounded in the first dimension).
We build a data structure of Lemma~\ref{lem-xue} on $\pi(\varPi')$ for $\mathcal{X} = \mathcal{P}$.
This gives us the desired data structure because the lightest pair in $\varPi' \Cap L$ for a query slab $L = [x^-,x^+] \times \mathbb{R}^{d-1}$ is $\phi$ iff the lightest pair in $\pi(\varPi') \Cap P$ for $P = [x^-,x^+] \times \mathbb{R}$ is $\pi(\phi)$.
We the consider the 2-box queries $\mathcal{B}_2$.
It suffices to handle the queries in the sub-collection $\mathcal{B}_2' \subseteq \mathcal{B}$ consisting of 2-boxes of the form $[x^-,\infty) \times [y^-,\infty) \times \mathbb{R}^{d-2}$.
We build a data structure of Lemma~\ref{lem-xue} on $\pi(\varPi')$ for $\mathcal{X} = \mathcal{Q}$.
Again, this gives us the desired data structure.

\subsection{Proof of Lemma~\ref{lem-candOnRd}}
The proof is the same as that of Lemma~\ref{lem-candOn}.
Let $\phi = \{a,b\} \in \varPi_2'$ where $\pi(a)$ is the southwest point of $\pi(\phi)$.
We claim that $b$ is the $\delta$-nearest-neighbor of among all the points whose $\pi$-images are to the northeast of $\pi(a)$ and colors are different from $\text{cl}(a)$.
To see this, suppose $B = [x^-,\infty) \times [y^-,\infty) \times \mathbb{R}^{d-2} \in \mathcal{B}_2'$ is a 2-box such that $\phi = \mathsf{CP}_\delta(\varPi_2 \Cap B)$.
Then $B$ contains all the points in $S$ whose $\pi$-images are to the northeast of $\pi(a)$.
Therefore, $\varPi_2 \Cap B$ contains all the bichromatic pairs consisting of $a$ and another point whose $\pi$-image is to the northeast of $a$.
As such, our claim holds.
Using this claim, if we charge every pair in $\phi \in \varPi_2'$ to its point whose $\pi$-image is the southwest point of $\pi(\phi)$, then every point in $S$ is charged at most once, which implies $|\varPi_2'| \leq n$.

\subsection{Proof of Lemma~\ref{lem-domcorrect}}
It is clear that $\varPhi \subseteq \mathsf{Bich}(S) \Cap B$, because every query we made above returns a bichromatic pair of points in $S \cap B$.
Thus, $\mathsf{CP}_\delta(\varPhi) \in \mathsf{Bich}(S) \Cap B$.
To see $|\mathsf{CP}_\delta(\varPhi)|_\delta \leq (1+\varepsilon) \cdot |\mathsf{CP}_\delta(\mathsf{Bich}(S) \Cap B)|_\delta$, let $\phi^* = \mathsf{CP}_\delta(\mathsf{Bich}(S) \Cap B)$.
It suffices to show $|\phi|_\delta \leq (1+\varepsilon) \cdot |\phi^*|_\delta$ for some $\phi \in \varPhi$.
If $\phi^* \subseteq S(\mathbf{u}_i)$ for some $i \in [t]$, then the pair $\phi$ returned by the data structure stored at $\mathbf{u}_i$ satisfies the desired property.
If $\phi^* \subseteq$ for some $i \in [t']$, then the pair $\phi$ returned by the data structure stored at $\mathbf{v}_i$ satisfies the desired property.
If $\phi^* \subseteq$ for some $i \in [t'']$, then the pair $\phi$ returned by the data structure stored at $\mathbf{w}_i$ satisfies the desired property.
Finally, if all of the above are not true, then $\phi^*$ must be $o$-anchored for some $o \in O$.
In this case, the pair $\phi$ returned by the anchored data structure $\mathbf{C}$ for the query $(B,o)$ satisfies the desired property.
As a result, $|\mathsf{CP}_\delta(\varPhi)|_\delta \leq (1+\varepsilon) \cdot |\mathsf{CP}_\delta(\mathsf{Bich}(S) \Cap B)|_\delta$.

\subsection{Proof of Theorem~\ref{thm-hard}}
For convenience, assume $n$ is even.
We first consider strip queries.
Construct $n/2$ red points $a_1,\dots,a_{n/2}$ where $a_i = (-i,n^2-in)$.
Construct $n/2$ blue points $b_1,\dots,b_{n/2}$ where $b_i = (i,in-n^2)$.
One can easily see that $L_2(a_i,b_j) \leq L_2(a_{i'},b_{j'})$ if $i' \geq i$ and $j' \geq j$.
We define $S_1 = \{a_1,\dots,a_{n/2},b_1,\dots,b_{n/2}\}$.
Then for all $i,j \in [n/2]$, the bichromatic pair $\{a_i,b_j\}$ is a candidate pair for strip queries.
Indeed, $\{a_i,b_j\}$ is the closest bichromatic pair of points in $S_1 \cap P$ for the strip $P = [-i,j] \times \mathbb{R}$, since $S_1 \cap P = \{a_1,\dots,a_i,b_1,\dots,b_j\}$.
Therefore, the number of the candidate pairs in $S_1$ for strip queries is $\Omega(n^2)$.
Next, we consider the quadrant queries.
Construct $n/2$ red points $a_1,\dots,a_{n/2}$ where $a_i = (-i,n^2-in)$.
Construct $n/2$ blue points $b_1,\dots,b_{n/2}$ where $b_i = (n^2-in,-i)$.
One can easily see that $L_2(a_i,b_j) \leq L_2(a_{i'},b_{j'})$ if $i' \geq i$ and $j' \geq j$.
We define $S_2 = \{a_1,\dots,a_{n/2},b_1,\dots,b_{n/2}\}$.
Then for all $i,j \in [n/2]$, the bichromatic pair $\{a_i,b_j\}$ is a candidate pair for quadrant queries.
Indeed, $\{a_i,b_j\}$ is the closest bichromatic pair of points in $S_2 \cap Q$ for the quadrant $Q = [-i,\infty) \times [-j,\infty)$, since $S_2 \cap Q = \{a_1,\dots,a_i,b_1,\dots,b_j\}$.
Therefore, the number of the candidate pairs in $S_2$ for quadrant queries is $\Omega(n^2)$.

\section{Well-behaved query spaces} \label{appx-well}
In this section, we give some examples of well-behaved query spaces, defined in Definition~\ref{def-well}.

Let $\mathcal{P}^\text{v}$ be the collection of vertical strips in $\mathbb{R}^2$.
We show that $\mathcal{P}^\text{v}$ is well-behaved on any set $\varPi$ of point pairs in $\mathbb{R}^2$.
We first verify the first condition in Definition~\ref{def-well}.
Consider a pair $\phi = \{a,b\} \in \varPi$.
Define $X_\phi = [\min\{\mathsf{x}(a),\mathsf{x}(b)\},\max\{\mathsf{x}(a),\mathsf{x}(b)\}] \times \mathbb{R} \in \mathcal{P}^\text{v}$.
Clearly, $X_\phi$ is the smallest vertical strip containing $\phi$ (i.e., $X_\phi \subseteq X$ for any $X \in \mathcal{P}^\text{v}$ containing $\phi$).
We then verify the second condition in Definition~\ref{def-well}.
Suppose $\phi = \{a,b\} \in \varPi$ and $\psi = \{a,c\} \in \varPi$ are strongly adjacent.
Then $a$ is a vertex of $\mathsf{BB}(\{a,b,c\})$, which implies that $a$ is either the leftmost point or the rightmost point in $\{a,b,c\}$.
Assume $a$ is the leftmost point in $\{a,b,c\}$ without loss of generality.
Then $X_\phi = [\mathsf{x}(a),\mathsf{x}(b)] \times \mathbb{R}$ and $X_\psi = [\mathsf{x}(a),\mathsf{x}(c)] \times \mathbb{R}$.
It follows that either $X_\phi \subseteq X_\psi$ or $X_\psi \subseteq X_\phi$.
Therefore, $\mathcal{P}^\text{v}$ is well behaved on $\varPi$.

Let $\mathcal{Q}^\text{ne}$ be the collection of northeast quadrants in $\mathbb{R}^2$.
We show that $\mathcal{P}^\text{v}$ is well-behaved on any set $\varPi$ of NW-SE pairs in $\mathbb{R}^2$.
We first verify the first condition in Definition~\ref{def-well}.
Consider a pair $\phi = \{a,b\} \in \varPi$.
Define $X_\phi = [\min\{\mathsf{x}(a),\mathsf{x}(b)\},\infty) \times [\min\{\mathsf{y}(a),\mathsf{y}(b)\},\infty) \in \mathcal{Q}^\text{ne}$.
Clearly, $X_\phi$ is the smallest northeast quadrant containing $\phi$ (i.e., $X_\phi \subseteq X$ for any $X \in \mathcal{Q}^\text{ne}$ containing $\phi$).
We then verify the second condition in Definition~\ref{def-well}.
Suppose $\phi = \{a,b\} \in \varPi$ and $\psi = \{a,c\} \in \varPi$ are strongly adjacent.
Then $a$ is a vertex of $\mathsf{BB}(\{a,b,c\})$.
Since $\phi$ and $\psi$ are both NW-SE, $a$ is either the northwest vertex or the southeast vertex of $\mathsf{BB}(\{a,b,c\})$.
Assume $a$ is the northwest vertex of $\mathsf{BB}(\{a,b,c\})$, without loss of generality.
Then $a$ is the northwest point of $\phi$ and also the northwest point of $\psi$.
As such, $X_\phi = [\mathsf{x}(a),\infty) \times [\mathsf{y}(b),\infty)$ and $X_\phi = [\mathsf{x}(a),\infty) \times [\mathsf{y}(c),\infty)$.
It follows that either $X_\phi \subseteq X_\psi$ or $X_\psi \subseteq X_\phi$.
Therefore, $\mathcal{Q}^\text{ne}$ is well behaved on $\varPi$.

Let $\mathcal{L}'$ be the collection of slabs in $\mathbb{R}^d$ of the form $[x^-,x^+] \times \mathbb{R}^{d-1}$.
Using the same argument as that for $\mathcal{P}^\text{v}$, we can prove that $\mathcal{L}'$ is well-behaved on any set $\varPi$ of point pairs in $\mathbb{R}^d$.

Let $\mathcal{B}_2'$ be the collection of 2-boxes in $\mathbb{R}^d$ of the form $[x^-,\infty) \times [y^-,\infty) \mathbb{R}^{d-2}$.
Define a map $\pi: \mathbb{R}^d \rightarrow \mathbb{R}^2$ as $(x_1,\dots,x_d) \mapsto (x_1,x_2)$.
Using the same argument as that for $\mathcal{Q}^\text{ne}$, we can prove that $\mathcal{B}_2'$ is well-behaved on any set $\varPi$ of point pairs in $\mathbb{R}^d$ satisfying that $\pi(\phi)$ is a NW-SE pair for all $\phi \in \varPi$.

\nocite{bkos_2000}
\nocite{afshani2009optimal}
\nocite{tao2004range_aggregate}
\nocite{guibas1983computing}
\nocite{abrahamsen2017range}
\nocite{arya2000approximate}
\nocite{oh2018approximate}
\nocite{corral2000closest}
\nocite{volpato2010fast}
\nocite{gupta1993algorithms}

\newpage
\bibliography{my_bib.bib}

\begin{thebibliography}{10}

\bibitem{abam2009power}
M.~A. Abam, P.~Carmi, M.~Farshi, and M.~Smid.
\newblock On the power of the semi-separated pair decomposition.
\newblock In {\em Workshop on Algorithms and Data Structures}, pages 1--12.
  Springer, 2009.

\bibitem{abrahamsen2017range}
Mikkel Abrahamsen, Mark de~Berg, Kevin Buchin, Mehran Mehr, and Ali~D Mehrabi.
\newblock Range-clustering queries.
\newblock {\em arXiv preprint arXiv:1705.06242}, 2017.

\bibitem{afshani2009optimal}
Peyman Afshani and Timothy~M Chan.
\newblock Optimal halfspace range reporting in three dimensions.
\newblock In {\em Proceedings of the twentieth Annual ACM-SIAM Symposium on
  Discrete Algorithms}, pages 180--186. Society for Industrial and Applied
  Mathematics, 2009.

\bibitem{agarwal1991euclidean}
Pankaj~K Agarwal, Herbert Edelsbrunner, Otfried Schwarzkopf, and Emo Welzl.
\newblock Euclidean minimum spanning trees and bichromatic closest pairs.
\newblock {\em Discrete \& Computational Geometry}, 6(3):407--422, 1991.

\bibitem{agarwal1999geometric:range_search}
Pankaj~K Agarwal, Jeff Erickson, et~al.
\newblock Geometric range searching and its relatives.
\newblock {\em Contemporary Mathematics}, 223:1--56, 1999.

\bibitem{agarwal2005geometric}
Pankaj~K Agarwal, Sariel Har-Peled, and Kasturi~R Varadarajan.
\newblock Geometric approximation via coresets.
\newblock {\em Combinatorial and computational geometry}, 52:1--30, 2005.

\bibitem{arya2000approximate}
Sunil Arya and David~M Mount.
\newblock Approximate range searching☆.
\newblock {\em Computational Geometry}, 17(3-4):135--152, 2000.

\bibitem{chazelle2008approximate}
Bernard Chazelle, Ding Liu, and Avner Magen.
\newblock Approximate range searching in higher dimension.
\newblock {\em Computational Geometry}, 39(1):24--29, 2008.

\bibitem{corral2000closest}
Antonio Corral, Yannis Manolopoulos, Yannis Theodoridis, and Michael
  Vassilakopoulos.
\newblock Closest pair queries in spatial databases.
\newblock In {\em ACM SIGMOD Record}, volume~29, pages 189--200. ACM, 2000.

\bibitem{bkos_2000}
M.~de~Berg, M.~Kreveld, M.~Overmars, and O.~Schwarzkopf.
\newblock {\em Computational Geometry}.
\newblock Springer Verlag, 2nd edition, 2000.

\bibitem{eppstein1995dynamic}
David Eppstein.
\newblock Dynamic euclidean minimum spanning trees and extrema of binary
  functions.
\newblock {\em Discrete \& Computational Geometry}, 13(1):111--122, 1995.

\bibitem{frahling2005coresets}
Gereon Frahling and Christian Sohler.
\newblock Coresets in dynamic geometric data streams.
\newblock In {\em Proceedings of the thirty-seventh annual ACM symposium on
  Theory of computing}, pages 209--217. ACM, 2005.

\bibitem{guibas1983computing}
L.~J. Guibas and J.~Stolfi.
\newblock On computing all north-east nearest neighbors in the
  $\textnormal{L}_1$-metric.
\newblock {\em Information Processing Letters}, 17(4):219--223, 1983.

\bibitem{gupta2006range}
Prosenjit Gupta.
\newblock Range-aggregate query problems involving geometric aggregation
  operations.
\newblock {\em Nordic journal of Computing}, 13(4):294--308, 2006.

\bibitem{gupta2014data}
Prosenjit Gupta, Ravi Janardan, Yokesh Kumar, and Michiel Smid.
\newblock Data structures for range-aggregate extent queries.
\newblock {\em Computational Geometry: Theory and Applications},
  2(47):329--347, 2014.

\bibitem{gupta1993algorithms}
Prosenjit Gupta, Ravi Janardan, and Michiel Smid.
\newblock Algorithms for some intersection searching problems involving curved
  objects.
\newblock 1993.

\bibitem{har2006coresets}
Sariel Har-Peled.
\newblock Coresets for discrete integration and clustering.
\newblock In {\em International Conference on Foundations of Software
  Technology and Theoretical Computer Science}, pages 33--44. Springer, 2006.

\bibitem{har2004coresets}
Sariel Har-Peled and Soham Mazumdar.
\newblock On coresets for k-means and k-median clustering.
\newblock In {\em Proceedings of the thirty-sixth annual ACM symposium on
  Theory of computing}, pages 291--300. ACM, 2004.

\bibitem{oh2018approximate}
Eunjin Oh and Hee-Kap Ahn.
\newblock Approximate range queries for clustering.
\newblock {\em arXiv preprint arXiv:1803.03978}, 2018.

\bibitem{shan2003spatial}
Jing Shan, Donghui Zhang, and Betty Salzberg.
\newblock On spatial-range closest-pair query.
\newblock In {\em International Symposium on Spatial and Temporal Databases},
  pages 252--269. Springer, 2003.

\bibitem{sharathkumar2007range}
R.~Sharathkumar and P.~Gupta.
\newblock Range-aggregate proximity queries.
\newblock {\em Technical Report IIIT/TR/2007/80. IIIT Hyderabad, Telangana},
  500032, 2007.

\bibitem{smid1995closest}
Michiel Smid.
\newblock {\em Closest point problems in computational geometry}.
\newblock Citeseer, 1995.

\bibitem{tao2004range_aggregate}
Yufei Tao and Dimitris Papadias.
\newblock Range aggregate processing in spatial databases.
\newblock {\em IEEE Transactions on Knowledge and Data Engineering},
  16(12):1555--1570, 2004.

\bibitem{volpato2010fast}
Nilton Volpato and Arnaldo Moura.
\newblock A fast quantum algorithm for the closest bichromatic pair problem,
  2010.

\bibitem{xue2018new}
Jie Xue, Yuan Li, Saladi Rahul, and Ravi Janardan.
\newblock New bounds for range closest-pair problems.
\newblock {\em Proceedings of the 34th International Symposium on Computational
  Geometry}, 2018.

\end{thebibliography}

\end{spacing}

\end{document}